\documentclass[fleqn,usenatbib]{mnras}

\usepackage[T1]{fontenc}
\usepackage{ae,aecompl}

\usepackage{graphicx}	
\usepackage{amsmath}	
\usepackage{amssymb}	
\graphicspath{images/}
\usepackage{xcolor}
\usepackage{float}
\usepackage{sidecap}
\usepackage{cleveref}

\newcommand{\swift}{Swift~J1749}
\newcommand{\nicer}{\textit{NICER}}
\newcommand{\cps}{\,cts\,s$^{-1}$}	
\newcommand{\csim}{$\sim$\,}

\newcommand{\chis}{$\chi^2$}
\newcommand{\NH}{$N_\mathrm{H}$}

\newcommand{\Mdot}{$\dot{M}$}

\title[Type-I X-ray Bursts from Swift~J1749]{Thermonuclear Type-I X-ray Bursts and Burst Oscillations from the Eclipsing AMXP Swift~J1749.4--2807}

\author[A. C. Albayati et al.]{
A.~C.~Albayati,$^{1}$\thanks{E-mail: a.c.albayati@soton.ac.uk}
P.~Bult,$^{2,3}$ 
D.~Altamirano,$^{1}$
J.~Chenevez,$^{4}$
S.~Guillot,$^{5}$\newauthor
\hspace{0.15cm}T.~G{\"u}ver,$^{6,7}$
G.~K.~Jaisawal,$^{4}$
C.~Malacaria,$^{8}$
G.~C.~Mancuso,$^{9,10}$
A.~Marino,$^{11,12,13}$\newauthor
\hspace{0.15cm}M.~Ng,$^{14}$
A.~Sanna,$^{15}$
T.~E.~Strohmayer$^{3}$
\vspace{0.2cm}
\\
$^{1}$School of Physics and Astronomy, University of Southampton, Southampton, SO17 1BJ, UK\\ 
$^{2}$Department of Astronomy, University of Maryland, College Park, MD 20742, USA\\
$^{3}$Astrophysics Science Division, NASA Goddard Space Flight Center, Greenbelt, MD 20771, USA\\ 
$^{4}$DTU Space, Technical University of Denmark, Elektrovej 327-328, Lyngby DK-2800, Denmark\\
$^{5}$Institut de Recherche en Astrophysique et Plan\'{e}tologie, UPS-OMP, CNRS, CNES, 9 avenue du Colonel Roche, BP 44346, F-31028 \\ \hspace{1mm}Toulouse Cedex 4, France\\
$^{6}$Istanbul University, Science Faculty, Department of Astronomy and Space Sciences, Beyaz\i t, 34119, \.Istanbul, Turkey \\
$^{7}$Istanbul University Observatory Research and Application Center, Istanbul University 34119, \.Istanbul Turkey\\
$^{8}$International Space Science Institute (ISSI), Hallerstrasse 6, 3012 Bern, Switzerland\\
$^{9}$Instituto Argentino de Radioastronom\'{\i}a (CCT-La Plata, CONICET; CICPBA), C.C. No. 5, 1894 Villa Elisa, Argentina\\
$^{10}$Facultad de Ciencias Astron\'omicas y Geof\'{\i}sicas, Universidad Nacional de La Plata, Paseo del Bosque s/n, 1900 La Plata, Argentina\\
$^{11}$Institute of Space Sciences (ICE, CSIC), Campus UAB, Carrer de Can Magrans s/n, E-08193 Barcelona, Spain\\
$^{12}$Institut d’Estudis Espacials de Catalunya (IEEC), Carrer Gran Capità 2–4, E-08034 Barcelona, Spain\\
$^{13}$INAF, Istituto di Astrofisica Spaziale e Fisica Cosmica, Via U. La Malfa 153, I-90146 Palermo, Italy\\
$^{14}$MIT Kavli Institute for Astrophysics and Space Research, Massachusetts Institute of Technology, 77 Massachusetts Avenue, \\ \hspace{3mm}Cambridge, MA 02139, USA\\
$^{15}$Dipartimento di Fisica, Universit `a degli Studi di Cagliari, SP Monserrato-Sestu km 0.7, Monserrato 09042, Italy
}

\date{Accepted XXX. Received YYY; in original form ZZZ}

\pubyear{2022}

\begin{document}
\label{firstpage}
\pagerange{\pageref{firstpage}--\pageref{lastpage}}
\maketitle

\begin{abstract}

Swift~J1749.4-2807 is the only known eclipsing accreting millisecond X-ray pulsar. 
In this paper, we report on 7 thermonuclear (Type-I) X-ray bursts observed by \nicer\ during its 2021 outburst. The first 6 bursts show slow rises and long decays, indicative of mixed H/He fuel, whereas the last burst shows fast rise and decay, suggesting He-rich fuel. 
Time-resolved spectroscopy of the bursts revealed typical phenomenology (i.e., an increase in black body temperature during the burst rise, and steady decrease in the decay), however they required a variable \NH. We found that the values of \NH\ during the bursts were roughly double those found in the fits of the persistent emission prior to each burst. We interpret this change in absorption as evidence of burst-disc interaction, which we observe due to the high inclination of the system. 
We searched for burst oscillations during each burst and detected a signal in the first burst at the known spin frequency of the neutron star (517.92\,Hz). This is the first time burst oscillations have been detected from Swift~J1749.4-2807. 
We further find that each X-ray burst occurs on top of an elevated persistent count rate. We performed time-resolved spectroscopy on the combined data of the bursts with sufficient statistics (i.e., the clearest examples of this phenomenon) and found that the black body parameters evolve to hotter temperatures closer to the onset of the bursts. We interpret this as a consequence of an unusual marginally stable burning process similar to that seen through mHz QPOs.

\vspace{4em}
\end{abstract}

\begin{keywords}
stars: neutron -- stars: individual (Swift~J1749.4--2807) -- X-rays: binaries -- X-rays: bursts\vspace{-1em}
\end{keywords}



\section{Introduction}

In a low-mass X-ray binary (LMXB) a neutron star or a black hole primary accretes matter from a stellar mass companion star via Roche lobe overflow. This material generates a vast amount of X-rays as it releases gravitational potential energy in the accretion disc around the primary \citep[see, e.g.,][for a review]{tau06}.
In the case of neutron star LMXBs matter is accreted onto the neutron star where it spreads across the surface and is compressed by the intense gravity. When the matter reaches a critical density, it ignites in a run-away thermonuclear explosion -- a Type-I X-ray burst (hereafter simply ``X-ray burst" or ``burst").
X-ray bursts appear as a sudden increase in X-ray emission over timescales of seconds, where the burst rises are typically $\lesssim$\,1 -- 10\,s, followed by a roughly exponential decay over tens to hundreds of seconds \citep[e.g.,][]{gal08}. 
Spectral fitting of X-ray bursts has revealed peak black body temperatures of 1 -- 3\,keV \citep[e.g., MINBAR\footnote{ \url{https://burst.sci.monash.edu/minbar}},][]{minbar}, and the ignited material can have a pure or mixed composition of hydrogen, helium, and sometimes carbon \citep[see, e.g.,][for reviews]{lew93, str03}. 
X-ray photons released by an X-ray burst can interact with the accretion disc, leading to changes in the disc structure, mass outflows from the disc, and a drag being exerted on the disc \citep[e.g.,][and references therein]{deg16, deg18}.

Swift\,J1749.4–2807 (hereafter \swift) was discovered in 2006 through observations made with the Neil Gehrels Swift Observatory \citep[\textit{Swift};][]{geh04} Burst Alert Telescope \citep[BAT;][]{kri13} as a burst-only source \citep[GRB060602B;][]{sch06, bea06}. Evidence from the soft spectra suggested that it might have been an X-ray burst rather than a classical gamma-ray burst \citep{pal06}. 
The burst origin was confirmed to be from an accreting neutron star in 2009 when another outburst from the source was observed with \textit{Swift}/BAT, \textit{Swift} X-Ray Telescope \citep[XRT;][]{bur03}, and \textit{XMM-Newton}'s European Photon Imaging Camera \citep[EPIC;][]{xmm}. The detection of an X-ray burst lasting \csim10\,s was reported during the 2009 outburst, and an approximate bolometric flux of $7^{+4}_{-2} \times10^{-8}$\,erg\,s$^{-1}$\,cm$^{-2}$ was estimated \citep{wij09}. The detection of this X-ray burst allowed for an upper limit estimation on the distance to the source of $6.7\pm1.3$\,kpc.
A persistent X-ray counterpart was found with the \textit{Swift}/XRT data, and supported by archival \textit{XMM-Newton} data showing a constant faint point source with coordinates consistent with the \textit{Swift}/XRT findings \citep{hal06, wij09}. 

\swift\ was detected again in 2010 by the \textit{INTEGRAL} Joint European X-Ray Monitor \citep[JEM-X2;][]{jemx} and observed further by \textit{Swift} \citep{pav10}, and the Rossi X-ray Timing Explorer \citep[\textit{RXTE};][]{jah06}. The observations from \textit{RXTE} revealed that \swift\ is an accreting millisecond X-ray pulsar \citep[AMXP, see][for reviews]{dis22, pat21}, detecting strong coherent pulsations at 517.9\,Hz and at its first over-tone \citep[1035.8\,Hz;][]{alt10, alt11, boz10}. The orbital period is 8.82\,hrs, suggesting a minimum companion mass of 0.475\,M$_{\odot}$ \citep{bel10, str10}. The \textit{RXTE} observations also revealed that the system was eclipsing \citep{mar10}, making it the first (and so far, only) detected eclipsing AMXP, allowing for the inclination of the system to be constrained to $i \approx 74^{\circ} - 77^{\circ}$ \citep{alt11}. During the 2010 outburst, the second X-ray burst from \swift\ was detected \citep{che10, fer11}, however statistics from the \textit{INTEGRAL}/JEM-X2 data were not sufficient to perform detailed spectral analysis.

After over 10 years in quiescence, a new outburst of \swift\ was detected by \textit{INTEGRAL}/JEM-X during Galactic bulge monitoring observations carried out on 28 February - 1 March 2021 \citep{2021ATel14427}. \textit{NICER} turned to the source on 1 March (MJD 59274) and observed regularly for the duration of the outburst \citep{2021ATel14428}. In this paper we report on the detection of 7 X-ray bursts with \textit{NICER}, and their detailed analysis. The detailed pulsar timing analysis of the full outburst has been presented in \citet{san22}, and analysis of the outflows and spectral evolution of this outburst has been presented in \citet{mar22}.

\begin{table}
\setlength{\tabcolsep}{5pt}
\centering
 \caption{Details of the \textit{NICER} Observation IDs analysed in this paper. Quoted exposure times are as after processing. Only the last digit (x) of ObsIDs are listed, where ObsID = 465801010x. Times are listed in Terrestrial Time. }
 \label{tab:ob}
 \begin{tabular}{cccccc}
  \hline
  ObsID &           & Start Time        & Date                      & Exposure              & X-ray  \\
    x   &           &\scriptsize{(MJD)} & \scriptsize{(DD-MM-YY)}   & \scriptsize{(sec)}    & Bursts \\
  \hline
  1     &           & 59274.61971       & 01-03-21                  & 9272                  & - \\
  2     &           & 59275.00551       & 02-03-21                  & 18150                 & Yes \\
  3     &           & 59276.03994       & 03-03-21                  & 9783                  & Yes \\
  4     &           & 59277.01551       & 04-03-21                  & 11637                 & Yes \\
  5     &           & 59278.04650       & 05-03-21                  & 16957                 & Yes \\
  6     &           & 59281.20661       & 08-03-21                  &  5263                 & - \\
  7     &           & 59282.04564       & 09-03-21                  &  11283                & Yes \\
  8     &           & 59283.01358       & 10-03-21                  &  12035                & - \\
  \hline
 \end{tabular}
\end{table}

\begin{figure*}	
    \centering
    \includegraphics[width=\linewidth, trim=0 0 0 0, clip]{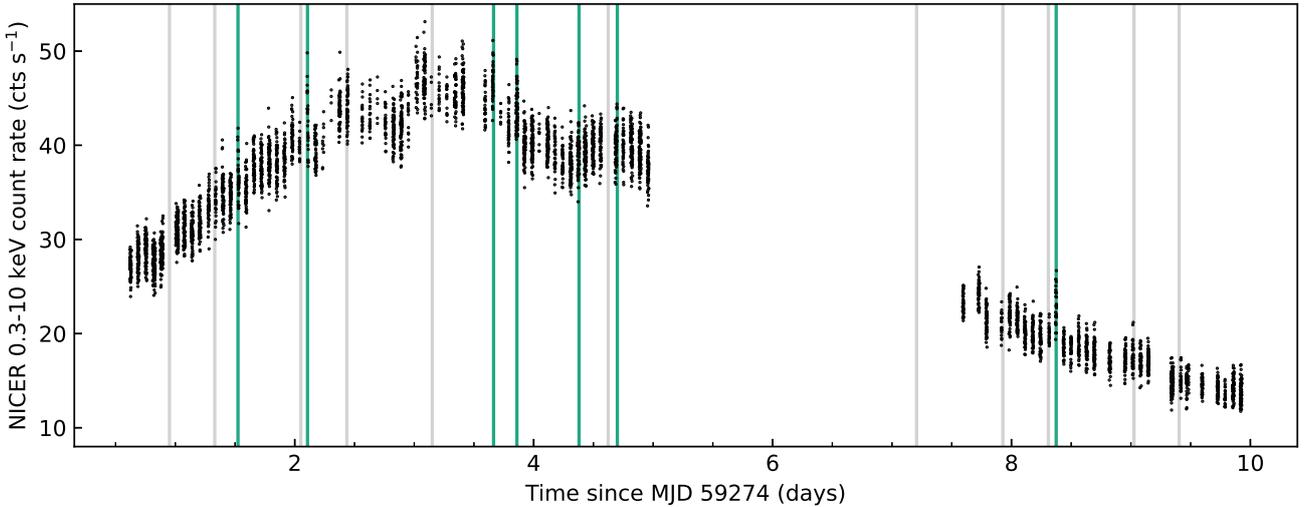}
    \caption{\textit{NICER} long-term light curve of the section of Swift~J1749's 2021 outburst which contains Type-I X-ray bursts with 16\,s time resolution in the 0.3 -- 10\,keV energy band. The outburst contains 7 thermonuclear X-ray bursts detected in March 2021, denoted by teal lines. 11 eclipses were also detected during this time, and are denoted by grey lines (note that there was a small section of eclipse-only data at \csim7.2\,days).
    }
    \label{fig:outburst}
\end{figure*}


\section{Observations}

\textit{NICER} observed \swift\ between 1 March and 1 May 2021, generating a total of 33 Observation IDs (ObsIDs 4658010101 -- 4658010133). We identified the section of the outburst containing X-ray bursts and selected 8 ObsIDs encompassing this section of the outburst to analyse in detail \citep[see][for analysis on a wider range of ObsIDs from this outburst]{san22,mar22}. These ObsIDs, each containing several data segments (or ``pointings")\footnote{After filtering, the 8 ObsIDs we analysed each contain between 6 -- 16 pointings, typically lasting \csim1000 -- 2000\,s.}, are listed in Table \ref{tab:ob}. 
To process the data, we used \texttt{HEASOFT} v6.28 \citep{heasoft} and \texttt{NICERDAS} v7 with standard filtering criteria. The total good exposure after processing was 94.4\,ksec.

\section{Data Analysis and Results}

\subsection{Outburst Evolution and Occurrence of X-ray Bursts}
\label{sec:outburst}

\begin{figure*}
    \centering
    \includegraphics[width=\linewidth, trim=0 0 0 0, clip]{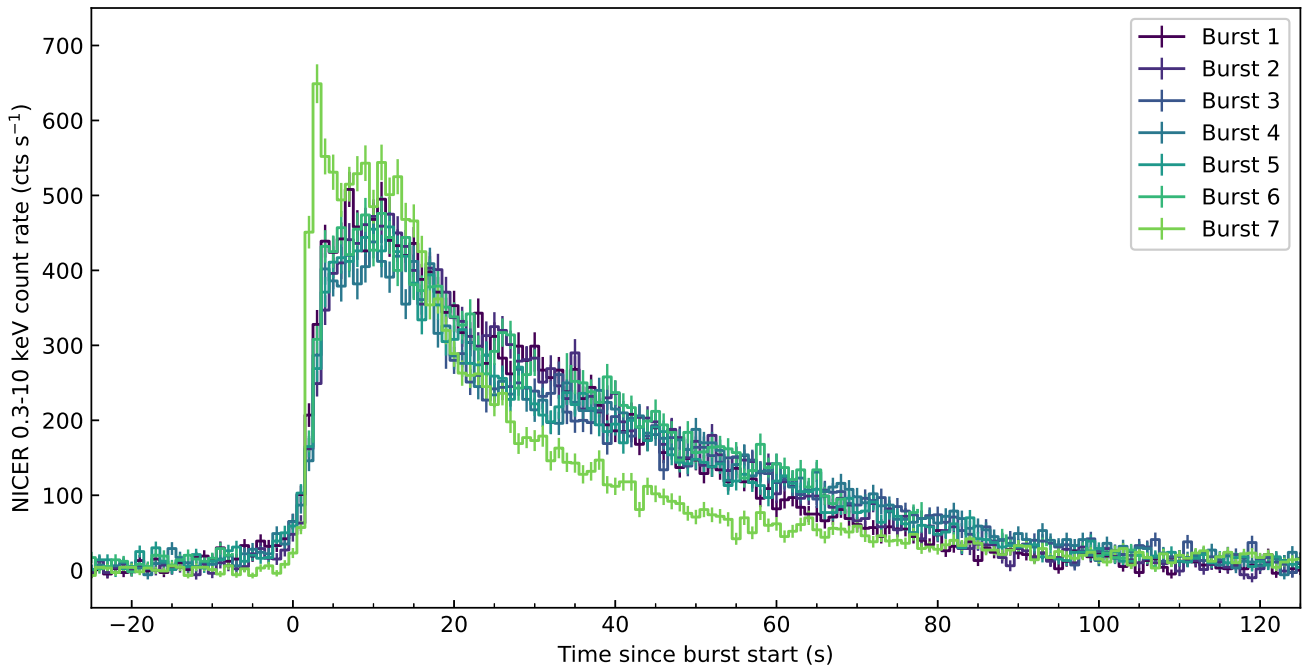}
    \caption{Background-subtracted light curves of the 7 X-ray bursts from Swift~J1749 aligned by burst onset (with manual optimisation for alignment in this figure) with 1\,s time resolution in the 0.3 -- 10\,keV energy band. The rise of B3 was not observed, so manual adjustments were made to optimise tail alignment.
     }
    \label{fig:bursts}
\end{figure*}

The outburst evolution of \swift\ between 1 -- 10 March 2021 (MJD 59274-59283), including X-ray bursts and eclipses, is shown in Figure \ref{fig:outburst}. 
We constructed a 0.3 -- 10\,keV long-term light curve using 16\,s bins. We searched for X-ray bursts and found 7 in total across 5 ObsIDs (see Table \ref{tab:ob}). In this paper, we do not analyse the full outburst as observed by \nicer, but only the ObsIDs where X-ray bursts were observed.

From MJD 59274.61971 ($t=0.6$\,days in Figure \ref{fig:outburst}, with reference time MJD 59274) the light curve shows the persistent count rate to rise from \csim28\cps\ to \csim36\cps\ over one day, at which point we detected the first X-ray burst (MJD 59275.52449). The second observed X-ray burst occurred half a day later (MJD 59276.10658) at a persistent rate of \csim42\cps. The persistent count rate peaked at approximately 50\cps\ at $t \approx 3.0$\,days, and began to drop until the beginning of a large visibility gap at $t \approx 5.0$\,days where it reached \csim37\cps. During this time, \nicer\ observed 4 more X-ray bursts (at MJD 59277.66427, 59277.85985, 59278.37967, 59278.70041) and 6 eclipses in total.
Observations resumed at $t \approx 7.2$\,days during an eclipse, where the persistent count rate was at \csim24\cps\ following this eclipse. We saw two more eclipses as the persistent count rate decreased further to roughly 20\cps\ at $t \approx 8.3$\,days, where a bright X-ray burst occurred (MJD 59282.37445). After the final X-ray burst observed by \nicer\ the persistent count rate continued to decrease. 

Inspection of the pointings containing X-ray bursts revealed that the count rate appears to increase before the onset of each X-ray burst by \csim20\%. We hereafter refer to this enhanced emission immediately prior to each X-ray burst as ``pre-burst emission".

\subsection{X-ray Bursts}

\subsubsection{Light Curves}
 
We extracted individual X-ray burst light curves in the 0.3 -- 10\,keV energy band using 1\,s bins (Figure \ref{fig:bursts}). We will refer to individual X-ray bursts as B1, B2, ... and B7 for the first, second, ... and seventh bursts respectively. 

In order to define the burst start and end times, first we identified the peak of the burst as the 1\,s bin with the highest count rate. We defined the start of each X-ray burst by searching backwards in time from the peak bin, and identifying the first time bin in which the count rate drops below 10\% of the peak. 
The burst rise time ($t_{rise}$) was defined as the time between the start and peak of the X-ray burst. We found the end of each X-ray burst by taking the median of 10\,s data segments from the burst peak, and searching forward in time for the first segment to fall below 10\% of the peak count rate. We defined the burst end to be the start time of this segment. The decay time ($t_{decay}$) is defined as the time between the peak and the end of the X-ray burst.

When considering the persistent count rate with regards to the background rate outside of the X-ray burst, in order to differentiate it from how we define persistent emission in our spectral analysis, we will hereafter refer to the persistent count rate as the ``continuum count rate". We found the continuum count rate by taking the median count rate of the data from the beginning of the pointing including an X-ray burst until 30\,s before the burst peak. There was no data collected due to a drop in telemetry during the rise of B3, so we manually adjusted the start time such that the tail of the burst was aligned with the other bursts, and calculated the continuum count rate by taking the median of the first 500\,s of data from the pointing containing the burst. For this reason, B3 is excluded from spectral analysis.

We find that the first 6 X-ray bursts in this outburst from \swift\ are all similar in profile, with an average rise time (excluding B3) of 9.7\,s ($\sigma = 3.9$\,s for the distribution of burst rise times, Figure \ref{fig:bursts}). This is longer than previously reported X-ray bursts from this source\footnote{We note that these bursts were detected by instruments with different energy bands and sensitivities to \nicer, \cite{wij09} detected by \textit{Swift}/BAT (>10\,keV) and \cite{fer11} by \textit{INTEGRAL}/JEM-X2 (3 -- 20\,keV).}. With a rise time of 2\,s, a more rapid decay and a higher peak count rate than the first 6 bursts, B7 more closely resembles the profiles of historic X-ray bursts from this source \cite[see,][]{fer11}. From these rise times, we can infer that the composition is likely H-rich fuel in the case of B1 -- 6, and comparatively He-rich for B7 \citep{gal08}.

Recent observations of X-ray bursts from SAX~J1808.4--3658 and MAXI~J1807+132 with \nicer\ data have revealed short plateaus, or ``pauses", in their rises \citep{bul19,alb21}. We looked at the 0.1\,s binned light curves to check for these, but found none.

\subsubsection{Persistent Emission}
\label{sec:persistent}

To fit the persistent emission, we took the pointing immediately before the one containing each X-ray burst, or the one before that if the one immediately before the X-ray burst was eclipsed\footnote{This decision was made as our tests had shown that the data immediately prior to the X-ray bursts, including the raised pre-burst emission, spectrally evolve.}. This pointing used for the persistent emission was within the same observation in all cases. Background and total spectra were created using \texttt{nibackgen3C50} \citep{nibackgen}, and the total spectra were rebinned using the optimal binning scheme of \citet{kaa16} with the additional requirement that each spectral bin contains at least 50 photons. In the case of B1, B2, B4, B5, and B7, we fitted the 1 -- 9\,keV (chosen due to background noise dominating outside this range) energy spectrum using the \texttt{Xspec} v.12.11.1 \citep{xspec} model

\begin{quote}
    \texttt{tbabs $\times$ powerlaw}\ ,
\end{quote}

\noindent setting the photoelectric cross-sections and the element abundances to the values provided by \cite{ver96} and \cite{wil00} respectively. We initially included a disc black body component, however we found that this component was not significant in all cases, so it was omitted. For all spectrally analysed X-ray bursts, we found an average column density of \NH\ $ = (3.52\,\pm\,0.03) \times10^{22}$\,cm$^{-2}$ 
which broadly agrees with values in literature \citep[e.g.][]{fer11, 2021ATel14428, mar22}. 

The photon indices, absorbed 0.5 -- 10\,keV fluxes \citep[following][]{mar22}, and reduced \chis\ obtained from modelling the persistent emissions are reported in Table \ref{tab:pe}. The errors are quoted at a 90\% confidence interval.

\begin{table}
\setlength{\tabcolsep}{3.75pt}
\centering
 \caption{The photon indices, absorbed 0.5 -- 10\,keV fluxes, and reduced \chis\ obtained from modelling the persistent emissions of \swift, with all errors quoted at a 90\% confidence interval.}
 \label{tab:pe}
 \begin{tabular}{ccccc}
  \hline
  Burst & \NH             & Photon    & Absorbed Flux                                       & $\chi^2_\nu$ \\
       & \scriptsize{($10^{22}$\,cm$^{-2}$)}  &Index     &\scriptsize{(10$^{-10}$\,erg\,cm$^{-2}$\,s$^{-1}$)}    & \scriptsize{($\chi^2/\mathrm{dof}$)}\\
  \hline
  1 & $3.59\pm0.08$          & $2.36\pm0.05$ & $2.56\pm0.02$ & $215/162$ \\
  2 & $3.62\pm0.08$          & $2.36\pm0.05$ & $3.10\pm0.03$ & $196/159$ \\
  4 & $3.63\pm0.09$          & $2.50\pm0.06$ & $3.05\pm0.03$ & $134/144$ \\
  5 & $3.53\pm0.06$          & $2.43\pm0.04$ & $2.82\pm0.02$ & $238/167$ \\
  6 & $3.55\pm0.07$          & $2.44\pm0.04$ & $2.99\pm0.02$ & $141/163$ \\
  7 & $3.11\pm0.09$          & $2.19\pm0.06$ & $1.49\pm0.02$ & $164/138$ \\
  \hline
 \end{tabular}
\end{table}

\subsubsection{Time-Resolved Spectroscopy of X-ray Bursts}
\label{sec:trs}

We used a 0.1\,s binned 1 -- 9\,keV light curve of each X-ray burst (using the previously calculated start and end times) to generate GTIs that set the temporal boundaries for the bins that we use for time-resolved spectroscopy (TRS). We combined 0.1\,s light curve bins until we had a time interval containing at least 1000 counts. The last bin was discarded as it usually had insufficient counts for spectral fitting. Individual spectra were extracted for each TRS bin, and were rebinned using optimal binning to a minimum of 25 counts per energy bin. A single background spectrum was extracted over the whole X-ray burst using \texttt{nibackgen3C50} and was used in fitting each spectra in the X-ray burst.

We fitted each TRS spectrum in the 1 -- 9\,keV energy band with the \texttt{Xspec} model 

\begin{quote}
    \texttt{tbabs $\times$ (bbodyrad + powerlaw)}
\end{quote}

\noindent allowing \texttt{tbabs} and \texttt{bbodyrad} to vary, with the powerlaw component fixed to the best fit parameters from the persistent emission fits (see Table \ref{tab:pe}). We found the unabsorbed bolometric flux by fitting with \texttt{cflux} in the 0.001--100\,keV band on \texttt{bbodyrad}. Errors were found using \texttt{Xspec}'s \texttt{chain} utility and are quoted for a 90\% confidence interval.

We chose to allow the absorption to vary as using a fixed-\NH\ black body method (with \NH\ fixed to the values found in the persistent emission, Table \ref{tab:pe}) in most cases resulted in high reduced \chis\ values, and implementing the $f_a$-method \citep{wor13,wor15} to scale the persistent emission did not improve the fits\footnote{We note that allowing \NH\ to vary also improved the fits when we used the data immediately prior to the burst (including the raised pre-burst emission) as our background spectra. }. We did, however, find that the fit could be improved by accounting for varying absorption throughout the burst. In order to confirm that varying absorption produced the best fits, we also varied the powerlaw photon index, however this produced unphysically high photon indices. We provide a more detailed discussion and comparison of fits in Appendix \ref{sec:app}.

\begin{table*}
\setlength{\tabcolsep}{7pt}
\renewcommand{\arraystretch}{1.5}
\centering
 \caption{Overview of X-ray burst information and parameters calculated from time-resolved spectroscopy (left to right): burst number, Observation ID where only the last digit (x) of ObsIDs are listed (ObsID = 465801010x), the peak unabsorbed bolometric flux, total bolometric burst fluence, burst rise time, burst decay time, characteristic timescale, and average hydrogen column density.}
 \label{tab:bursts}
 \begin{tabular}{ccccccccc}
  \hline
  \#    & ObsID & Onset Time        & Peak $F_{\mathrm{bol}}$  & Fluence  & $t_{rise}$ & $t_{decay}$ & $\tau$ & Average \NH\\
        &       &\scriptsize{(MJD)} & \scriptsize{($10^{-9}$ erg\,s$^{-1}$\,cm$^{-2}$)}    & \scriptsize{($10^{-7}$ erg\,cm$^{-2}$)}   & \scriptsize{(s)}  & \scriptsize{(s)} & \scriptsize{(s)} &\scriptsize{(10$^{22}$\,cm$^{-2}$)}\\
  \hline
  1     & 2     & 59275.52449       & $15.85^{+3.91}_{-2.91}$      & $3.62\pm0.13$      & 8                 & 65                 & $22.84\pm4.08$ &    $5.63\pm0.17$ \\
  2     & 3     & 59276.10658       & $14.40^{+3.27}_{-2.43}$      & $3.71\pm0.13$      & 13                & 65                 & $25.79\pm4.19$ &    $5.51\pm0.16$ \\
  4     & 4     & 59277.85985       & $14.96^{+4.41}_{-3.05}$      & $3.55\pm0.13$      & 11                & 75                 & $23.70\pm5.03$ &    $5.52\pm0.14$ \\
  5     & 5     & 59278.37967       & $12.62^{+2.79}_{-2.13}$      & $3.36\pm0.12$      & 14                & 67                 & $26.66\pm4.21$ &    $5.60\pm0.17$ \\
  6     & 5     & 59278.70041       & $13.21^{+2.76}_{-2.11}$      & $3.49\pm0.13$      & 10                & 58                 & $26.43\pm3.90$ &    $5.46\pm0.17$ \\
  7     & 7     & 59282.37445       & $49.12^{+20.36}_{-14.06}$    & $3.84\pm0.35$      & 2                 & 50                 & $7.81\pm2.02$ &     $5.33\pm0.24$ \\
  \hline
 \end{tabular}
\end{table*}

\begin{figure*}
    \centering
    \includegraphics[width=\linewidth, trim=45 10 70 50, clip]{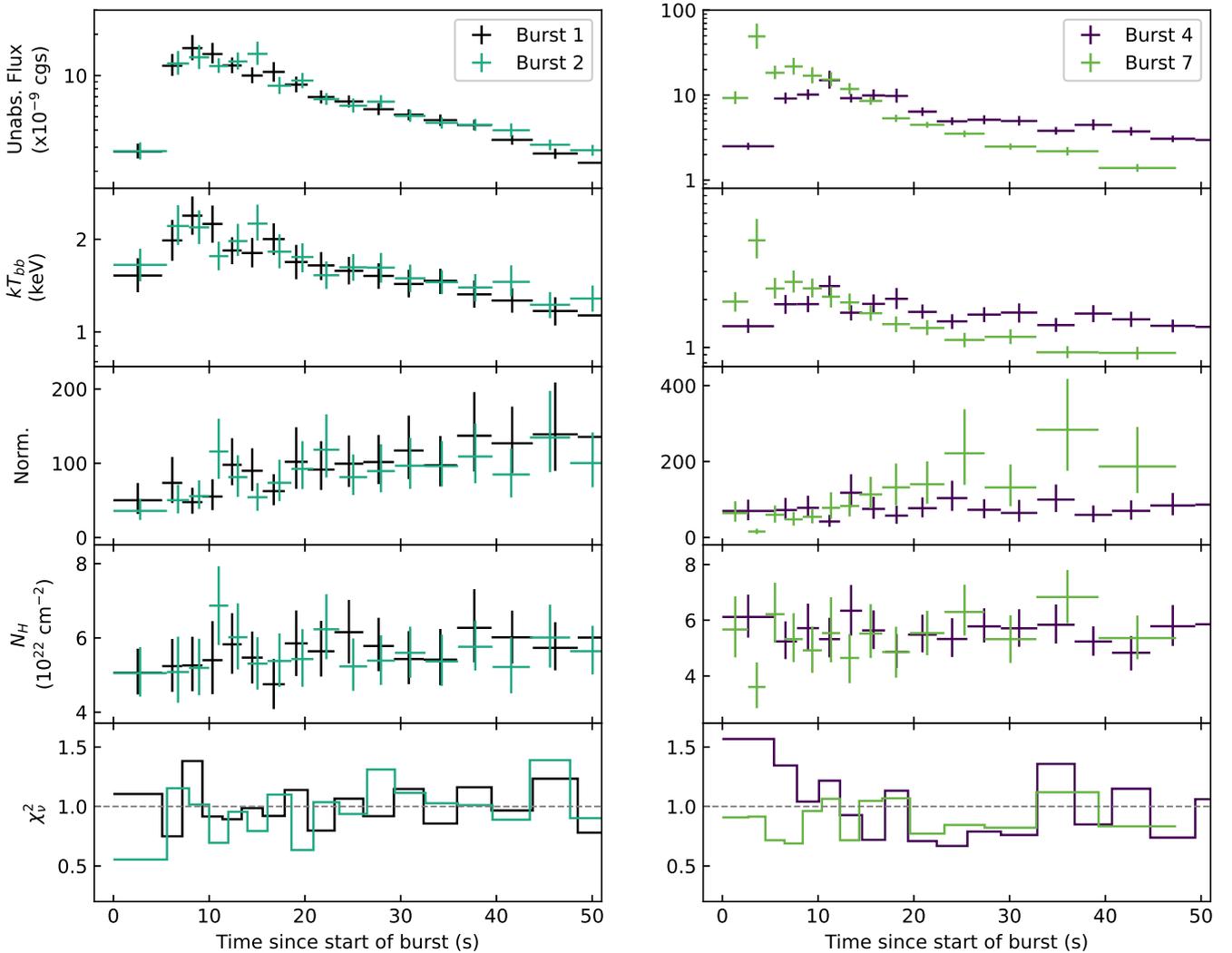}

    \caption{Time-resolved spectroscopy of bursts 1 and 2 (left), and 4 and 7 (right) using a variable \NH\ method. From top, we show the unabsorbed bolometric flux (in units of erg\,s$^{-1}$\,cm$^{-2}$), the black body temperature and normalisation, the variable Hydrogen column density and the reduced \chis. Note the log scaling in the first two rows.}
    \label{fig:trs}
\end{figure*}

In Figure \ref{fig:trs} we show the best-fit parameters for the time-resolved spectroscopy of B1, B2, B4, and B7 as examples of our results.
The bolometric unabsorbed fluxes ($F_{\mathrm{bol}}$) and black body temperatures ($kT_{bb}$) of all spectrally analysed bursts approximately follow the light curve contours, increasing during the burst rise and decreasing during the decay. This is expected for X-ray bursts \citep[see, e.g.,][for a review]{lew93}. The peak fluxes are reported in Table \ref{tab:bursts}.  
In Table \ref{tab:bursts}, we also report the X-ray burst fluences and time scales $\tau$ ($=$ fluence / peak $F_{bol}$), where the fluences were calculated over the burst duration (i.e., $t_{rise} \ + \ t_{decay}$).

We found that the values of \NH\ during the X-ray bursts were higher by an average of ($1.98\pm0.08$)\,$\times$10$^{22}$\,cm$^{-2}$ than those found in the fits of the persistent emission. 
From Figure \ref{fig:trs}, we can see that there was no clear evolution of \NH\ in any of the bursts, however they appear to be correlated with black body normalisation. This is likely due to poor statistics.

We note that none of these bursts exhibit photospheric radius expansion \citep[PRE; see, e.g.,][for a review]{gal08} characterised by an increase in black body normalisation and decrease in $kT_{bb}$ at the burst peak. The brightest X-ray burst in our data set (B7) is not as bright as those seen in historic data, therefore a new upper limit on distance is not possible to attain.


\subsection{Raised Pre-Burst Emission}

\begin{figure}
    \centering
    \includegraphics[width=\linewidth, trim=0 0 0 0, clip
    ]{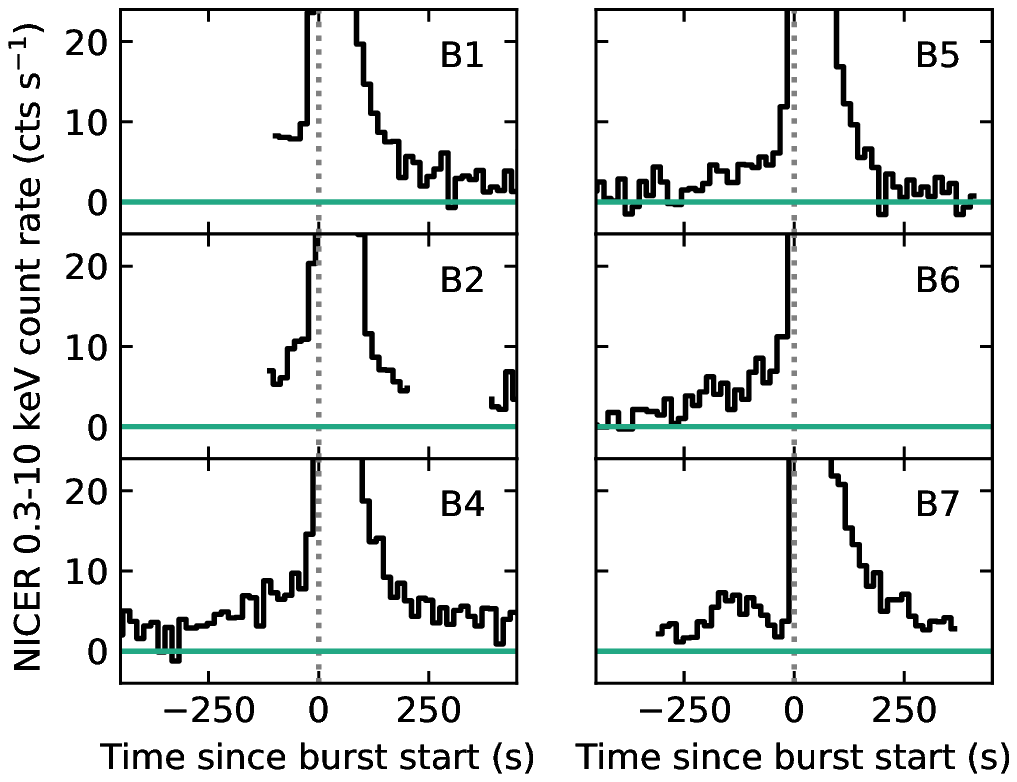}
    \caption{Baseline count rate subtracted light curves of the X-ray bursts (excluding B3) from Swift~J1749 with 16\,s time resolution in the 0.3 -- 10\,keV energy band. The burst start times are marked by vertical dotted lines. Error bars omitted for clarity, but are typically $\pm 1$\cps.     }
    \label{fig:elev_emis}
\end{figure}

\subsubsection{Light Curves of Pre-Burst Emission}

 To investigate the enhanced pre-burst emission in the X-ray burst light curves, we defined a ``baseline" count rate for each X-ray burst by taking the mean count rates of the pointings immediately before and after the pointing containing the X-ray burst, and taking an average of these two values. Baseline-subtracted light curves of each X-ray burst are shown in Figure \ref{fig:elev_emis}. The count rate before and after the bursts appear to be elevated compared with neighbor pointings.
 
 In the case of B1, B2, B4, B5, and B6 the pre-burst count rate increases by \csim8\cps\ before the burst onset. B7's pre-burst count rate also increases to \csim8\cps, but it dips back down to the baseline count rate \csim40\,s before the X-ray burst onset. In the light curves where we have sufficient data (B4-7), we can see that these phenomena occur from \csim200-250\,s before the burst onset.

\subsubsection{Time-Resolved Spectroscopy of Pre-Burst Emission}
\label{sec:pre-trs}

To investigate the spectral evolution of the emission leading up to the X-ray bursts, we considered the best examples of X-ray bursts with raised pre-burst count rate (B4, B5, and B6) and performed time-resolved spectroscopy. 

\begin{figure}
    \centering
    \includegraphics[width=\linewidth, trim=15 10 10 0, clip]{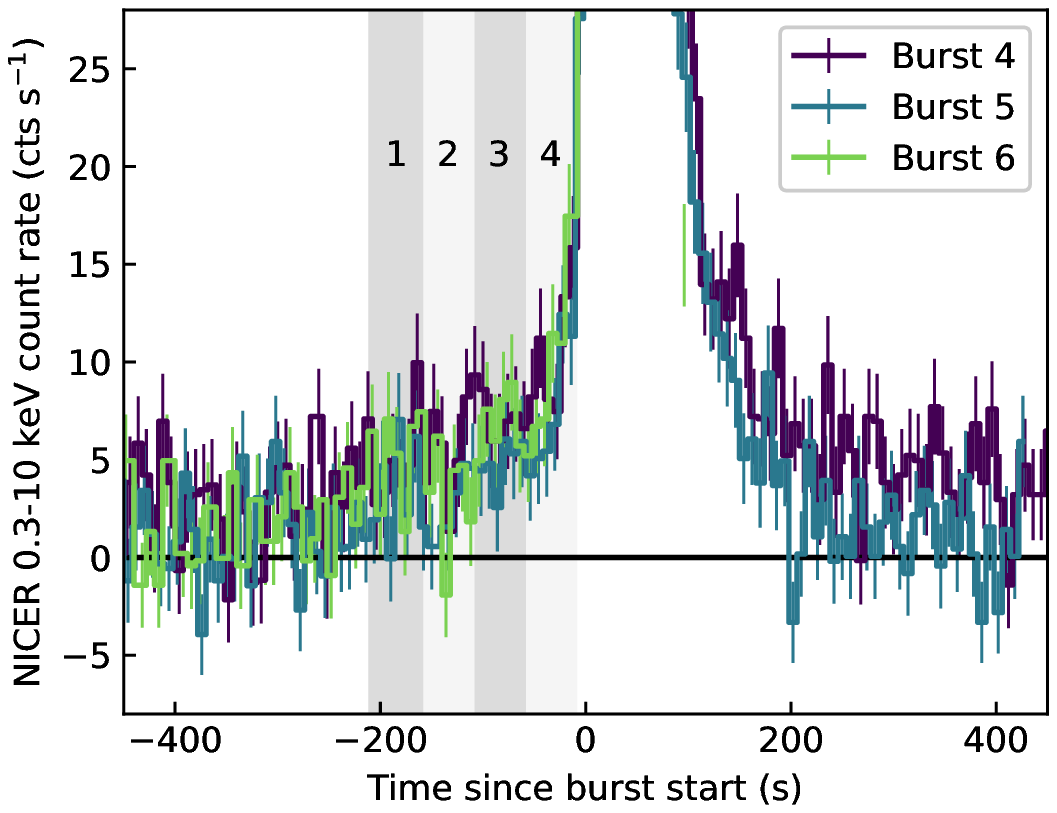}
    \caption{Baseline emission subtracted light curves of B4, B5, and B6 from Swift~J1749 with 8\,s time resolution in the 0.3 -- 10\,keV energy band. The 4 bins of 50\,s duration each used in the spectral analysis of the pre-burst rising persistent count rate are denoted by grey bands.
    }
    \label{fig:pre-spec_lc}
\end{figure}

We isolated 4 bins of data each 50\,s in length leading up to 10\,s before the start of the X-ray burst (see Figure \ref{fig:pre-spec_lc}). We extracted event files that combined all bin 1 segments, all bin 2 segments, etc., and the same for the persistent emission (data selection for persistent spectra defined in \S\ref{sec:persistent}). We justify combining the data in this way by noting that the persistent emission spectral shapes are consistent within errors, suggesting that all three observations sample the same source state.

We extracted background and total spectra in the 1--9\,keV energy band for each bin and the persistent emission using \texttt{nibackgen3C50}, and re-binned the total spectra using optimal binning to a minimum of 50\,cts per energy bin.  
While investigating the combined persistent emission spectrum, we found an absorption line at \csim7\,keV which we fitted with a Gaussian. This absorption line was also found by \cite{mar22} who identified it as a slightly blueshifted Fe~XXVI absorption line commonly found in high-inclination LMXBs \citep[see, e.g.,][]{pon14}. 

\begin{figure}
    \centering
    \includegraphics[width=\linewidth, trim=10 10 5 0, clip]{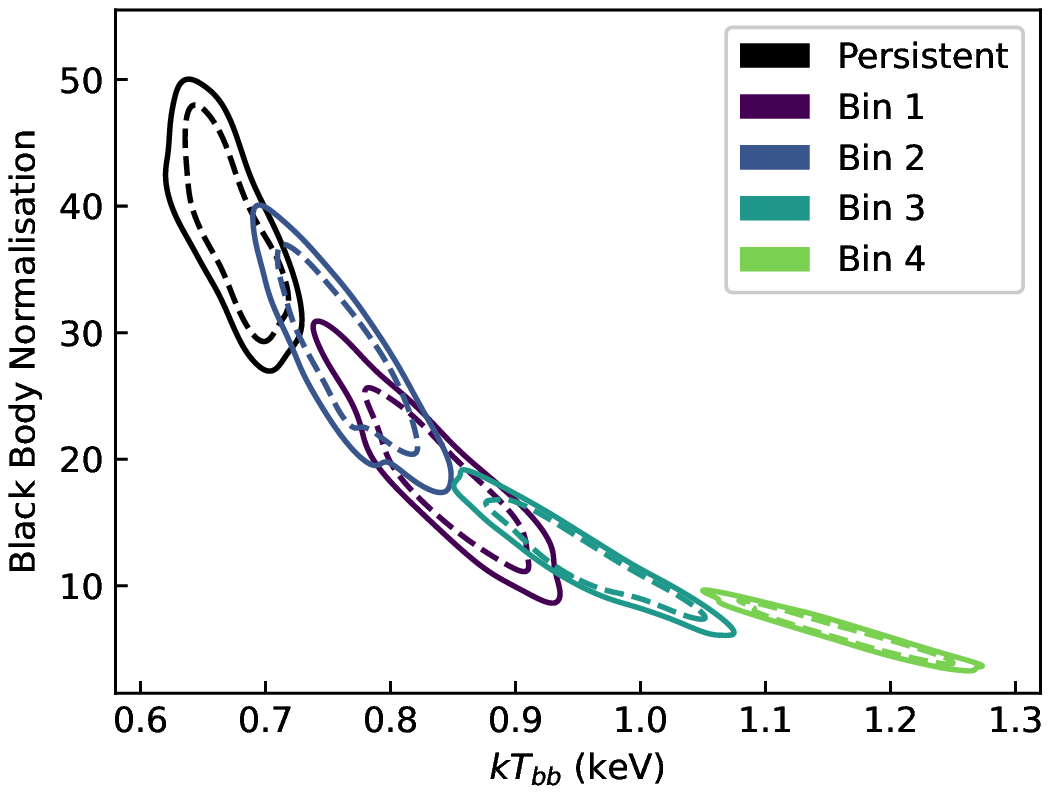}
    
    \caption{A contour plot created using MCMC, showing the black body parameter evolution from fitting the pre-burst emission with the model \texttt{tbabs $\times$ (diskbb + powerlaw)}, where \textit{only} \texttt{diskbb} was allowed to vary freely. Confidence levels are shown at 68\% (dotted line) and 90\% (solid line).}

    \label{fig:contours}
\end{figure}

We simultaneously fit all the bins and persistent spectra with the spectral model stated in \S\ref{sec:persistent}, including a Gaussian component to account for the absorption line found in the persistent spectrum. For these fits, we estimate the confidence regions using MCMC contours. Firstly, we fit the data allowing both the black body and power law parameters to vary freely, whilst the \NH\ parameters were tied. 
The black body temperature evolved to hotter temperatures compared with the persistent spectrum, whilst the power law parameters remained consistent with the persistent fits. 
In order to better constrain the black body parameters, we performed fits where the power law parameters were tied together for all the bins including the persistent spectrum, and only the black body parameters were allowed to vary freely. We show the MCMC contours from this fit in Figure \ref{fig:contours}. 
Compared with the persistent spectrum, the black body temperature ($kT_{bb}$) evolves towards hotter temperatures closer to the burst onset, whilst the normalisation decreases. 
We calculated the absorbed flux for each bin including the persistent bin, and found that it increases closer to the burst onset.

\subsection{Search for Burst Oscillations}

We searched all 7 X-ray bursts for the presence of burst oscillations using
a $Z_2^2$-based search method \citep{Buccheri1983}. First we corrected the
event times to the solar system barycentre using the DE-430 ephemeris
\citep{Folkner2014} and the source coordinates reported by \citet{Jonker2013}.
Next, we used the binary ephemeris reported by \citet{san22} to correct the
event times for the Doppler shift introduced by the orbital motion of the
neutron star.
For a given X-ray burst, we then used all events in the $0.5-10$\,keV energy range
within a $100$\,s time interval starting at the burst onset. We then used
a sliding window method to extract a series of views on the data, where we used
window durations of $T=4$, $8$, and $16$\,s with steps of $T/4$. For each
extracted window, we evaluated the $Z_2^2$ score on a grid of frequencies,
defined as a $4$ times oversampled Fourier frequency grid that was centred on
the $518$\,Hz neutron star spin frequency and has a width of 10\,Hz. If we assume that
Poisson counting statistics hold, then the $Z_2^2$ scores in absence of
a signal should follow a \chis\ distribution with four degrees of freedom.
Hence, if a measured $Z_2^2$ score exceeded the trial-adjusted $3\sigma$
detection threshold set by this distribution, then we adopted it as a burst
oscillation candidate.

A single candidate burst oscillation was found in B1, making it the first burst oscillation candidate found in an X-ray burst from \swift. The candidate had a
frequency of $517.92$\,Hz, which is entirely consistent with the known spin frequency of the neutron
star \citep{san22}. This candidate was found in both $8$\,s and $16$\,s duration
windows, with peak scores of $33$ and $40$, respectively. In both cases the
highest score was found when the window position was centred on $t=8$\,s,
(relative to the onset time, see the top panel of Figure \ref{fig:burst
oscillation}), meaning that the burst oscillation candidate was found at the
peak intensity of the X-ray burst. In order to verify the detection
significance of this candidate, we used a Monte Carlo approach to simulate the
true noise distribution of the search method \citep{bil19, Bult2021a}.
Compared to this simulated distribution, we find that the measured signal had
a detection significance of $3.7\sigma$. 

\begin{figure}
  \centering
  \includegraphics[width=\linewidth]{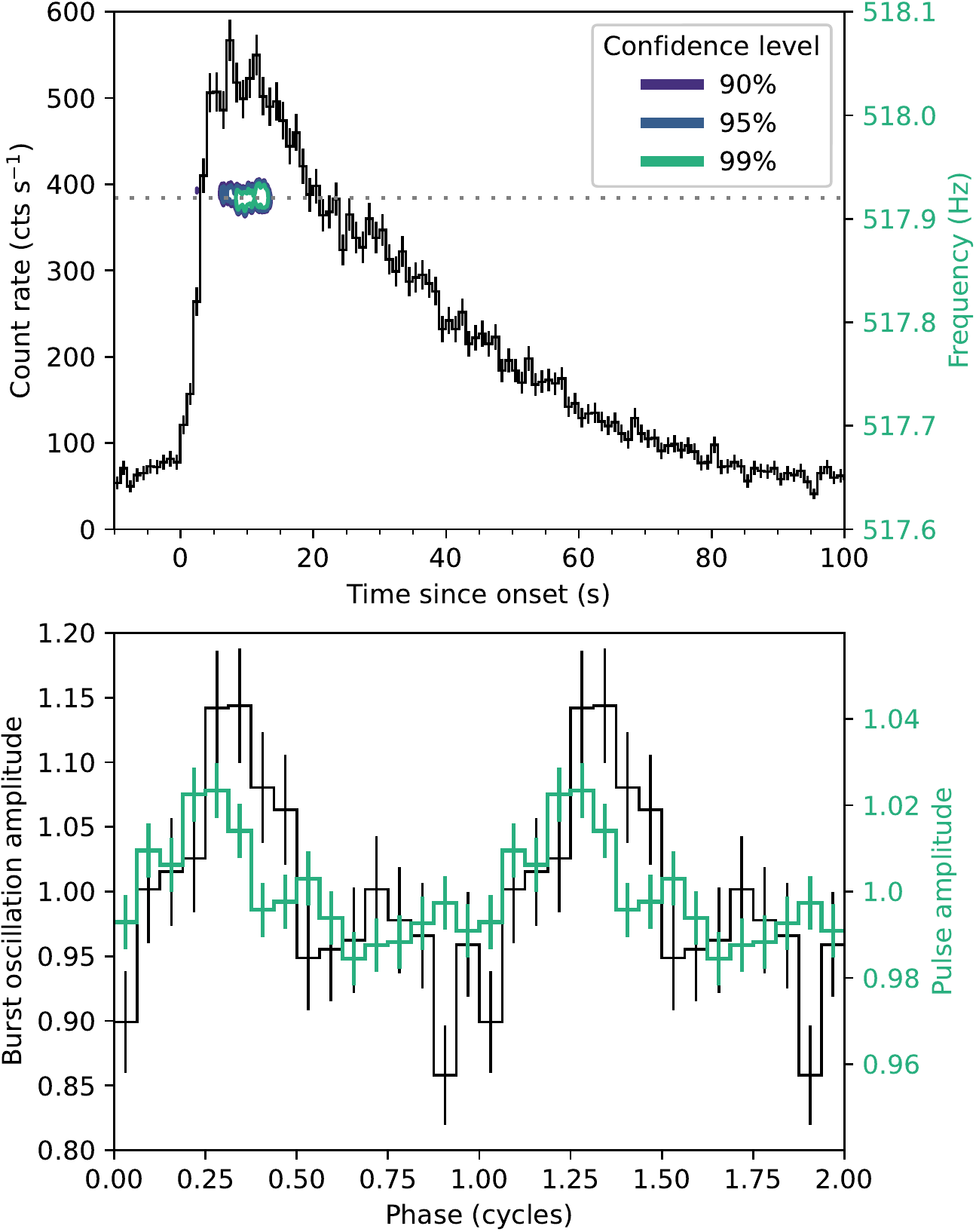}
  \caption{%
    Burst oscillation detection in Burst 1. The top panel shows the X-ray burst
    light curve at 0.5\,s time resolution (left axis) along with a dynamic
    power spectrum calculated using a $16$\,s duration sliding window (contours,
    right axis). The dotted line indicates the known 517.92\,Hz pulsar spin
    frequency. The bottom panel shows two cycles of the burst oscillation
    waveform (black) compared to the contemporaneous coherent pulsation in the
    non-burst emission \citep[teal,][]{san22}. Because the coherent
    pulsation has a much smaller amplitude, we plotted this second waveform
    against the zoomed-in scale of the left axis. 
  } 
  \label{fig:burst oscillation}
\end{figure}

Taking the epoch during which the burst oscillation is detected ($t=2-22$\,s after onset), we folded the data on the burst oscillation frequency to obtain
its waveform (Figure \ref{fig:burst oscillation}, bottom panel). This waveform
is somewhat asymmetric and has a fractional root mean square (rms) amplitude of $(7.1\pm1.5)\%$.
In an attempt to investigate the energy dependence of the burst oscillation
amplitude, we subdivided the energy range into three bands ($0.5-2$, $2-4$,
$4-10$\,keV), and constructed the burst oscillation waveform in each band. The
oscillation was significantly detected in each band, showing a consistent
phase and amplitude. Hence, no significant energy dependence could be detected.

Like accretion-powered, coherent pulsations from this
source, the burst oscillation has an asymmetric profile, pointing to
substantial harmonic content. Indeed, if we decompose the waveform into
harmonic components, we find a fundamental amplitude of $5.9\%$ rms and
a second harmonic amplitude of $3.4\%$ rms. These amplitudes are well in excess
of the amplitudes measured contemporaneously from the coherent pulsations
(Figure \ref{fig:burst oscillation}, bottom panel). Furthermore, the burst
oscillation appears to lag behind the coherent pulsation by about $0.07$
cycles. We note, however, that the phase of the coherent pulsation shows
a much larger scatter over the course of the outburst \citep{san22}, so we
interpret the burst oscillation as being in phase alignment with the coherent
pulsation, at least within the statistical uncertainty.

\section{Discussion}

In this paper we reported an analysis of 7 Type-I X-ray bursts observed by \nicer\ during the 2021 outburst of the AMXP \swift. 
From time-resolved spectroscopy we found enhanced \NH\ values during the X-ray bursts compared with those found in the persistent emission prior to each burst. 
We detected a burst oscillation signal during the peak of the first burst entirely consistent with the known spin frequency of the neutron star (517.92\,Hz). 
Additionally, we found that each X-ray burst occurs on top of an elevated persistent count rate. Time-resolved spectroscopy of the pre-burst emission showed that the black body parameters evolve to hotter temperatures closer to the onset of the X-ray burst. 
In this section we discuss each of these findings.

\subsection{Type-I X-ray Bursts} 
\label{sec:bursts}

\subsubsection{X-ray Burst Spectral Characteristics}

\begin{figure}
  \centering
  \includegraphics[width=\linewidth, trim=60 20 120 50, clip]{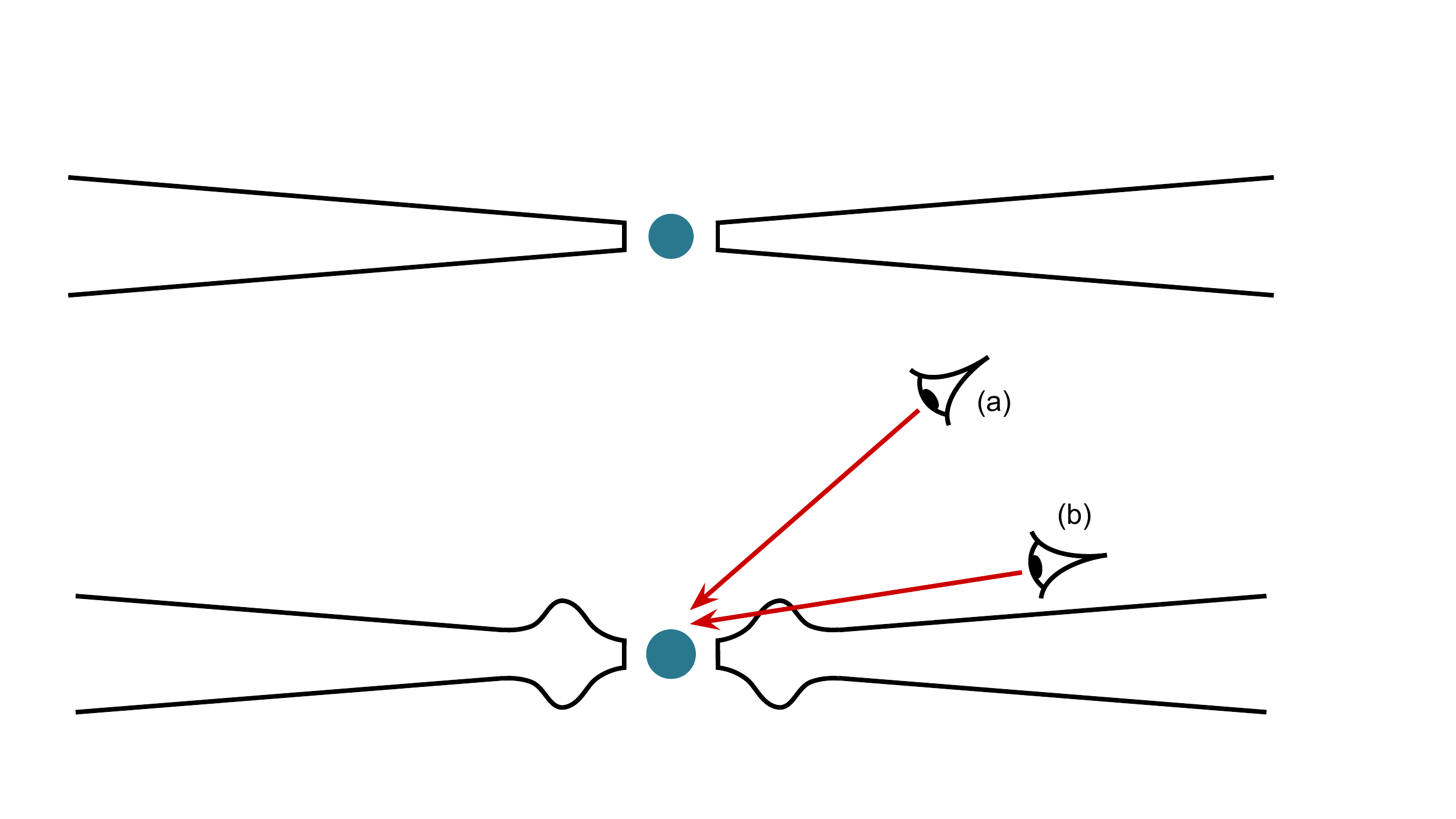}
  \caption{%
   Schematic of neutron star and accretion disc, viewed edge-on. Top: a standard thin disc, bottom: a scenario where a Type-I X-ray burst has caused the inner accretion disc to ``puff up". We can see that this change in disc geometry will not affect line-of-sight absorption in a low-inclination system (a), but will in a high-inclination system (b).  
  } 
  \label{fig:NH_schematic}
\end{figure}

We find that all X-ray bursts from \swift\ for which we performed time-resolved spectroscopy exhibit typical spectral phenomenology (an increase in black body temperature during the burst rise, and steady decrease in the decay). This did, however, require all the fits to be performed while allowing \NH\ to vary. \cite{mar22} also found that allowing \NH\ to vary during spectral analysis of the full 2021 outburst of \swift\ improves the spectral fits. However, \cite{mar22} does not report values as large as those obtained from our burst fitting.
We find that, during the X-ray bursts, the average \NH\ values were significantly higher than those found in the fits of the persistent emission. We interpret this raised \NH\ during the bursts as being a consequence of irradiation of the accretion disc by the X-ray burst emission. This interaction would cause the disc to ``puff up" during the X-ray burst, thus increasing the line-of-sight absorption, as illustrated in Figure \ref{fig:NH_schematic}. This interpretation is strengthened by the fact that we would only be able to observe this increased line-of-sight absorption if the system is high-inclination, as \swift\ is \citep[$i \approx 74^{\circ} - 77^{\circ}$;][]{alt11}. If our interpretation is correct, this would explain why these results have not, to our knowledge, been found in the time-resolved spectroscopy of other Type-I X-ray bursts (in low-inclination systems).

Allowing \NH\ to vary during the time-resolved spectroscopy of X-ray bursts has not, to our knowledge, been used to analyse X-ray bursts from eclipsing systems \citep[there are 8 known eclipsing X-ray bursters out of a total of 85 bursting sources;][]{minbar}. However, the varying \NH\ method has
been performed before on superbursts by \cite{bal04} and \cite{ kee14}\footnote{Interestingly, both \cite{bal04} and \cite{kee14} also find the Fe~XXVI absorption line discussed in \S\ref{sec:pre-trs}. We note that the sources discussed in both these papers are not high inclination (i.e. $i < 70^{\circ}$).}. Both studies found that \NH\ increases by an order of magnitude during the superburst compared with the persistent emission, and follows a clear evolution. \cite{bal04} explain the phenomenon with an increase in the scale height of the accretion disc, while \cite{kee14} suggest that disc wind driven by the superburst provides additional absorbing material. Whilst we see no clear evolution in \NH\ during the bursts themselves (likely due to poor statistics), the fact that we observe changes in \NH\ similar to those observed in much brighter superbursts is evidence that we might be seeing burst-disc interaction even during standard Type-I X-ray bursts. 

There are arguments for certain orbital phases being associated with higher absorption, for example, due to the accretion stream over and under-flowing the accretion disc \citep[see, e.g.,][]{mis16}. We checked to see if each burst was occurring at the same orbital phase, however the bursts all occurred at different points in the orbital phase.

We report the X-ray burst peak bolometric fluxes, fluences and $\tau$ values in Table \ref{tab:bursts} and will now compare them with the general population. In order to compare peak flux and fluence with \cite{gal08}, we calculated the normalised peak flux (defined as peak flux divided by Eddington flux) and normalised fluence ($U_b$, defined as fluence divided by Eddington flux), using Eddington flux found by \cite{kuu03}. We find that the normalised peak flux values for B1 -- 6, excluding B3, are all \csim0.2, placing them in the centre of the positively skewed Gaussian distribution of normalised peak fluxes for non-PRE bursts. B7 has a normalised peak flux of \csim0.7, placing it in the upper tail of the distribution. We find that all our analysed bursts have $U_b \approx 5$, placing them around the centre of the Gaussian distribution for non-PRE bursts. We also find that the values of $\tau$ for B1 -- 6, excluding B3, are very high (\csim23 -- 27\,s), placing them in the second peak in the bimodal distribution of non-PRE bursts. These values are in-line with what we expect due to the low peak fluxes of the bursts. For B7, which has a significantly higher peak flux than the other bursts, we find $\tau$ to be \csim8\,s, which is in the centre of the first peak of the bimodal distribution of the general population. These results are consistent with mixed H/He unstable burning \citep[see, e.g.,][]{gal17} where the \textit{rp}-process is dominant in B1 -- 6, while B7 is more helium-rich.

\subsubsection{Burst Oscillations}

We have, for the first time, detected a burst oscillation in an X-ray burst of
\swift. The burst oscillation was found in the peak of B1 at a frequency of
$517.92$\,Hz, which matches the known spin frequency of the neutron
star \citep{san22}. \cite{minbar} lists 16 sources which exhibit burst oscillations, three of which are persistent pulsars \citep[see also][]{bil19}. Here we report on the fourth persistent pulsar from which burst oscillations have been detected. These burst oscillations are typical with respect to fractional rms amplitude: the general population of burst oscillations have a median rms amplitude of 5\% \citep[][]{gal08,oot17}. We find burst oscillations in the peak of an X-ray burst, the burst phase in which 54\% of oscillations are found \citep{gal08}. However, only 25\% of bursts with burst oscillations have $\tau > 10$\,s, placing B1 in this minority ($\tau_{\mathrm B1} = 30$\,s). These results show that the burst oscillations we find in \swift\ are consistent with the general population.

\subsection{Raised Pre-Burst Emission}

We find that the persistent count rate slowly and systematically increases before the ignition of (almost) every X-ray burst we detected. This phenomenon has, to our knowledge, not been previously observed in other X-ray burst sources. From our spectral analysis, we find that the raised pre-burst emission is due to an increase in black body temperature of the thermal component in our model. Although we see that a thermal component is responsible, the statistical quality of the data did not allow us to resolve the spectrum into multiple thermal components representing disc emission and surface emission separately. Therefore, we cannot distinguish between disc and surface emission. 

If we assume that the increase in pre-burst emission is due to thermal disc emission, this implies that the mass accretion rate (\Mdot) through the disc rises in the \csim200\,s prior to the burst. One (unlikely) possibility is that the fluctuation in \Mdot\ is not intrinsically related to the burst. Instead these fluctuations might be stochastic variations that happen throughout the outburst and are by chance aligned with some of our bursts. The fact that the pre-burst light curve profile in B7 varies from the previous bursts suggests that the variations are just stochastic. We investigated this idea by searching for random rises in persistent count rate \textit{without} the onset of an X-ray burst\footnote{This was found by taking the median and maximum deviation in each pointing, and taking the absolute difference between the two. We visually inspected pointings showing values higher than \csim4\cps. Other than in pointings containing either a burst or eclipse, there were no systematic deviations. }.
The elevated emission has a clearly discernible pattern in the light curve with a deviation of \csim8\cps. Inspecting the light curve of the whole outburst we found no instance of such a deviation other than the ones found immediately preceding the X-ray bursts. We therefore conclude that the raised pre-burst emission is connected to the burst ignition.

We must also ask why the count rate, and hence \Mdot\, returns to normal after the X-ray burst occurred. A second possibility relating the raised pre-burst emission to the disc is that the X-ray burst itself acts as a feedback control on \Mdot. From the time-resolved spectroscopy we found that \NH\ increases during the burst, which we argue is due to the disc puffing up in response to irradiation by the burst. This change in disc geometry will affect the rate of mass flow from the disc onto the neutron star surface that the disc can support, so we could speculate that it is the radiative impact on the disc that forces it back into its lower \Mdot\ state.

The problem with this idea, however, is that there is no reason to believe that an increase in \Mdot\ should always trigger an X-ray burst. The extra flux contributed during the raised pre-burst phase is only a very small fraction of the accretion fluence between bursts, so it is exceedingly unlikely that an X-ray burst would ignite every time precisely \csim200\,s after \Mdot\ starts to increase. We therefore argue that the raised pre-burst emission is associated with the neutron star surface and not the accretion disc.

If we assume that the increase in pre-burst emission and rise in black body temperature is due to the stellar surface, we could attribute it to nuclear burning processes that lead into the unstable burst ignition. There are two phenomena discussed in the literature that are known to precede X-ray bursts: precursors and mHz quasi-periodic oscillations (mHz QPOs). 
``Precursors'' are short, burst-like events lasting a few seconds which have been interpreted as being due to, for example, (i) deep layer or multi-layer burst ignition \citep{bha07}, (ii) mechanisms related to multi-peaked bursts \citep{jon04}, or (iii) shock-generated X-ray bursts before superbursts \citep{kee11, kee12}. 
These short events described as precursors are not similar to the pre-burst features we see from \swift, which are over 200\,s and seem to be due to changes in the persistent emission rather than additional features on top of the persistent emission. Therefore we can dismiss that the raised pre-burst emission we see in our bursts is related to the same phenomena.

The second phenomena seen preceding X-ray bursts are mHz QPOs. \citet{rev01} discovered a class of low-frequency (7 -- 9\,mHz), soft energy (1 -- 5\,keV) QPO with fractional rms amplitudes of $\approx$\,2\%, which disappear after X-ray bursts. \citet{heg07} explained that these mHz QPOs could be caused by marginally stable nuclear burning on the neutron star's surface, where the burning is only oscillatory close to the luminosity boundary between stable and unstable burning. \citet{heg07} also shows that temperature fluctuations are possible in the burning layer during the oscillations. \citet{alt08} found that the triggering of a thermonuclear X-ray burst is also linked to the frequency of the QPO dropping below $\lesssim9$\,mHz. 
A possible interpretation of the pre-burst rises we see in our data is that these rises are the start of mHz QPOs that trigger an X-ray burst after just the rise of the first period.
If we consider the ``bump" before B7 to be half an oscillation, this gives us a possible QPO frequency of \csim2\,mHz, which is below the critical frequency found by \cite{alt08}. We also find a fractional rms amplitude of \csim25\%.
Whilst mHz QPOs are similar to what we observe in \swift, these values do not align with those established for typically seen mHz QPOs. Therefore we suggest that the raised pre-burst emission is related to mHz QPOs in that it is caused by a special case of marginally stable or confined burning process on the stellar surface that culminates in the unstable ignition of an X-ray burst. 

If our interpretation is correct, and the raised pre-burst emission is due to a special marginally stable burning process, then we should also consider why this phenomenon is not routinely observed in other sources. We posit two possible explanations. Firstly, we note that mHz QPOs are quite rare \citep[seen in only 6 sources;][and references therein]{man21}. Both observations and theory suggest that marginally stable burning only occur in a narrow range of luminosities \citep[$L_{2-20\,keV} \approx$ 5 -- 11$\,\times 10^{36}$\,erg s$^{-1}$;][]{alt08}, a range which mHz QPO sources tend to only pass through during their outbursts. \swift\ stays within a narrow range of luminosities ($L_{2-20\,keV}\approx $\ 1 -- 2$\,\times 10^{36}$\,erg s$^{-1}$) during its outburst. It is possible that this special marginally stable burning process occurs in this narrow range of luminosities that \swift\ happens to stay in during its outburst. Hence, specific neutron star and \Mdot\ parameters may simply place \swift\ in a part of the marginally stable burning process regime that is not sampled by other X-ray burst sources. 
Secondly, as a pulsar, \swift\ has unusually high harmonic content (i.e., the overtone is often stronger than the fundamental), pointing to a rather unusual inner accretion geometry \citep{alt11, san22}. If we assume that other classical mHz QPO sources are indeed non-pulsating sources (and that the lack of observed pulsations is not due to rotational axis alignment with the magnetic axis), \swift\ would be the first mHz QPO source that is also an accreting millisecond pulsar and so has a stronger magnetic field than non-pulsating sources. Hence, we could speculate that we are observing a mHz QPO that is either magnetically confined to a local region of the stellar surface and/or distorted by the viewing angle (where the high inclination of the system may play a role).


\section*{Acknowledgements}

A.A. thanks the referee for making helpful suggestions that improved the paper. A.A. also thanks M. \& L. Albayati for their support. D.A. acknowledges support from the Royal Society. S.G. acknowledges the support of the CNES. T.G. has been supported in part by the Turkish Republic, Presidency of Strategy and Budget project, 2016K121370. G.C.M. was partially supported by PIP 0113 (CONICET) and received financial support from PICT-2017-2865 (ANPCyT). A.M. is supported by the H2020 ERC Consolidator Grant “MAGNESIA” under grant agreement No. 817661 (PI: Rea) and National Spanish grant PGC2018-095512-BI00. A.M. is also partially supported by the program Unidad de Excelencia Maria de Maeztu CEX2020-001058-M, and by the PHAROS COST Action (No. CA16214).


\section*{Data Availability}

This work made use of data provided by the High Energy Astrophysics Science Archive Research Center (\textsc{heasarc}).




\bibliographystyle{mnras}
\bibliography{ref} 

\begin{thebibliography}{}
\makeatletter
\relax
\def\mn@urlcharsother{\let\do\@makeother \do\$\do\&\do\#\do\^\do\_\do\%\do\~}
\def\mn@doi{\begingroup\mn@urlcharsother \@ifnextchar [ {\mn@doi@}
  {\mn@doi@[]}}
\def\mn@doi@[#1]#2{\def\@tempa{#1}\ifx\@tempa\@empty \href
  {http://dx.doi.org/#2} {doi:#2}\else \href {http://dx.doi.org/#2} {#1}\fi
  \endgroup}
\def\mn@eprint#1#2{\mn@eprint@#1:#2::\@nil}
\def\mn@eprint@arXiv#1{\href {http://arxiv.org/abs/#1} {{\tt arXiv:#1}}}
\def\mn@eprint@dblp#1{\href {http://dblp.uni-trier.de/rec/bibtex/#1.xml}
  {dblp:#1}}
\def\mn@eprint@#1:#2:#3:#4\@nil{\def\@tempa {#1}\def\@tempb {#2}\def\@tempc
  {#3}\ifx \@tempc \@empty \let \@tempc \@tempb \let \@tempb \@tempa \fi \ifx
  \@tempb \@empty \def\@tempb {arXiv}\fi \@ifundefined
  {mn@eprint@\@tempb}{\@tempb:\@tempc}{\expandafter \expandafter \csname
  mn@eprint@\@tempb\endcsname \expandafter{\@tempc}}}

\bibitem[\protect\citeauthoryear{{Albayati} et~al.,}{{Albayati}
  et~al.}{2021}]{alb21}
{Albayati} A.~C.,  et~al., 2021, \mn@doi [\mnras] {10.1093/mnras/staa3657},
  \href {https://ui.adsabs.harvard.edu/abs/2021MNRAS.501..261A} {501, 261}

\bibitem[\protect\citeauthoryear{{Altamirano}, {van der Klis}, {Wijnands}  \&
  {Cumming}}{{Altamirano} et~al.}{2008}]{alt08}
{Altamirano} D.,  {van der Klis} M.,  {Wijnands} R.,   {Cumming} A.,  2008,
  \mn@doi [\apjl] {10.1086/527355}, \href
  {https://ui.adsabs.harvard.edu/abs/2008ApJ...673L..35A} {673, L35}

\bibitem[\protect\citeauthoryear{{Altamirano} et~al.,}{{Altamirano}
  et~al.}{2010}]{alt10}
{Altamirano} D.,  et~al., 2010, The Astronomer's Telegram, \href
  {https://ui.adsabs.harvard.edu/abs/2010ATel.2565....1A} {2565, 1}

\bibitem[\protect\citeauthoryear{{Altamirano} et~al.,}{{Altamirano}
  et~al.}{2011}]{alt11}
{Altamirano} D.,  et~al., 2011, \mn@doi [\apjl] {10.1088/2041-8205/727/1/L18},
  \href {https://ui.adsabs.harvard.edu/abs/2011ApJ...727L..18A} {727, L18}

\bibitem[\protect\citeauthoryear{{Arnaud}}{{Arnaud}}{1996}]{xspec}
{Arnaud} K.~A.,  1996, in {Jacoby} G.~H.,  {Barnes} J.,  eds,  Astronomical
  Society of the Pacific Conference Series Vol. 101, Astronomical Data Analysis
  Software and Systems V. p.~17

\bibitem[\protect\citeauthoryear{{Ballantyne} \& {Strohmayer}}{{Ballantyne} \&
  {Strohmayer}}{2004}]{bal04}
{Ballantyne} D.~R.,  {Strohmayer} T.~E.,  2004, \mn@doi [\apjl]
  {10.1086/382703}, \href
  {https://ui.adsabs.harvard.edu/abs/2004ApJ...602L.105B} {602, L105}

\bibitem[\protect\citeauthoryear{{Beardmore}, {Godet}  \&
  {Sakamoto}}{{Beardmore} et~al.}{2006}]{bea06}
{Beardmore} A.~P.,  {Godet} O.,   {Sakamoto} T.,  2006, GRB Coordinates
  Network, \href {https://ui.adsabs.harvard.edu/abs/2006GCN..5209....1B} {5209,
  1}

\bibitem[\protect\citeauthoryear{{Belloni}, {Stella}, {Bozzo}, {Israel}  \&
  {Campana}}{{Belloni} et~al.}{2010}]{bel10}
{Belloni} T.,  {Stella} L.,  {Bozzo} E.,  {Israel} G.,   {Campana} S.,  2010,
  The Astronomer's Telegram, \href
  {https://ui.adsabs.harvard.edu/abs/2010ATel.2568....1B} {2568, 1}

\bibitem[\protect\citeauthoryear{{Bhattacharyya} \&
  {Strohmayer}}{{Bhattacharyya} \& {Strohmayer}}{2007}]{bha07}
{Bhattacharyya} S.,  {Strohmayer} T.~E.,  2007, \mn@doi [\apj]
  {10.1086/510359}, \href
  {https://ui.adsabs.harvard.edu/abs/2007ApJ...656..414B} {656, 414}

\bibitem[\protect\citeauthoryear{{Bilous} \& {Watts}}{{Bilous} \&
  {Watts}}{2019}]{bil19}
{Bilous} A.~V.,  {Watts} A.~L.,  2019, \mn@doi [\apjs]
  {10.3847/1538-4365/ab2fe1}, \href
  {https://ui.adsabs.harvard.edu/abs/2019ApJS..245...19B} {245, 19}

\bibitem[\protect\citeauthoryear{{Bozzo}, {Belloni}, {Israel}  \&
  {Stella}}{{Bozzo} et~al.}{2010}]{boz10}
{Bozzo} E.,  {Belloni} T.,  {Israel} G.,   {Stella} L.,  2010, The Astronomer's
  Telegram, \href {https://ui.adsabs.harvard.edu/abs/2010ATel.2567....1B}
  {2567, 1}

\bibitem[\protect\citeauthoryear{{Buccheri} et~al.,}{{Buccheri}
  et~al.}{1983}]{Buccheri1983}
{Buccheri} R.,  et~al., 1983, \aap, \href
  {https://ui.adsabs.harvard.edu/abs/1983A&A...128..245B} {128, 245}

\bibitem[\protect\citeauthoryear{{Bult} et~al.,}{{Bult} et~al.}{2019}]{bul19}
{Bult} P.,  et~al., 2019, \mn@doi [\apjl] {10.3847/2041-8213/ab4ae1}, \href
  {https://ui.adsabs.harvard.edu/abs/2019ApJ...885L...1B} {885, L1}

\bibitem[\protect\citeauthoryear{{Bult} et~al.,}{{Bult}
  et~al.}{2021a}]{Bult2021a}
{Bult} P.,  et~al., 2021a, \mn@doi [\apj] {10.3847/1538-4357/abd54b}, \href
  {https://ui.adsabs.harvard.edu/abs/2021ApJ...907...79B} {907, 79}

\bibitem[\protect\citeauthoryear{{Bult} et~al.,}{{Bult}
  et~al.}{2021b}]{2021ATel14428}
{Bult} P.~M.,  et~al., 2021b, The Astronomer's Telegram, \href
  {https://ui.adsabs.harvard.edu/abs/2021ATel14428....1B} {14428, 1}

\bibitem[\protect\citeauthoryear{{Burrows} et~al.,}{{Burrows}
  et~al.}{2003}]{bur03}
{Burrows} D.~N.,  et~al., 2003, in \procspie. pp 1320--1325,
  \mn@doi{10.1117/12.461279}

\bibitem[\protect\citeauthoryear{{Chenevez} et~al.,}{{Chenevez}
  et~al.}{2010}]{che10}
{Chenevez} J.,  et~al., 2010, The Astronomer's Telegram, \href
  {https://ui.adsabs.harvard.edu/abs/2010ATel.2561....1C} {2561, 1}

\bibitem[\protect\citeauthoryear{{Degenaar}, {Koljonen}, {Chakrabarty}, {Kara},
  {Altamirano}, {Miller}  \& {Fabian}}{{Degenaar} et~al.}{2016}]{deg16}
{Degenaar} N.,  {Koljonen} K.~I.~I.,  {Chakrabarty} D.,  {Kara} E.,
  {Altamirano} D.,  {Miller} J.~M.,   {Fabian} A.~C.,  2016, \mn@doi [\mnras]
  {10.1093/mnras/stv2965}, \href
  {https://ui.adsabs.harvard.edu/abs/2016MNRAS.456.4256D} {456, 4256}

\bibitem[\protect\citeauthoryear{{Degenaar} et~al.,}{{Degenaar}
  et~al.}{2018}]{deg18}
{Degenaar} N.,  et~al., 2018, \mn@doi [\ssr] {10.1007/s11214-017-0448-3}, \href
  {https://ui.adsabs.harvard.edu/abs/2018SSRv..214...15D} {214, 15}

\bibitem[\protect\citeauthoryear{{Di Salvo} \& {Sanna}}{{Di Salvo} \&
  {Sanna}}{2022}]{dis22}
{Di Salvo} T.,  {Sanna} A.,  2022, in {Bhattacharyya} S.,  {Papitto} A.,
  {Bhattacharya} D.,  eds,  Astrophysics and Space Science Library Vol. 465,
  Astrophysics and Space Science Library. pp 87--124,
  \mn@doi{10.1007/978-3-030-85198-9_4}

\bibitem[\protect\citeauthoryear{{Ferrigno} et~al.,}{{Ferrigno}
  et~al.}{2011}]{fer11}
{Ferrigno} C.,  et~al., 2011, \mn@doi [\aap] {10.1051/0004-6361/201015033},
  \href {https://ui.adsabs.harvard.edu/abs/2011A&A...525A..48F} {525, A48}

\bibitem[\protect\citeauthoryear{{Folkner}, {Williams}, {Boggs}, {Park}  \&
  {Kuchynka}}{{Folkner} et~al.}{2014}]{Folkner2014}
{Folkner} W.~M.,  {Williams} J.~G.,  {Boggs} D.~H.,  {Park} R.~S.,   {Kuchynka}
  P.,  2014, Interplanetary Network Progress Report, \href
  {https://ui.adsabs.harvard.edu/abs/2014IPNPR.196C...1F} {42-196, 1}

\bibitem[\protect\citeauthoryear{{Galloway} \& {Keek}}{{Galloway} \&
  {Keek}}{2021}]{gal17}
{Galloway} D.~K.,  {Keek} L.,  2021, in {Belloni} T.~M.,  {M{\'e}ndez} M.,
  {Zhang} C.,  eds,  Astrophysics and Space Science Library Vol. 461, Timing
  Neutron Stars: Pulsations, Oscillations and Explosions. pp 209--262
  (\mn@eprint {arXiv} {1712.06227}), \mn@doi{10.1007/978-3-662-62110-3_5}

\bibitem[\protect\citeauthoryear{{Galloway} et~al.}{{Galloway}
  et~al.}{2008}]{gal08}
{Galloway} D.~K.,  et~al., 2008, \mn@doi [\apjs] {10.1086/592044}, \href
  {https://ui.adsabs.harvard.edu/abs/2008ApJS..179..360G} {179, 360}

\bibitem[\protect\citeauthoryear{{Galloway} et~al.,}{{Galloway}
  et~al.}{2020}]{minbar}
{Galloway} D.~K.,  et~al., 2020, \mn@doi [\apjs] {10.3847/1538-4365/ab9f2e},
  \href {https://ui.adsabs.harvard.edu/abs/2020ApJS..249...32G} {249, 32}

\bibitem[\protect\citeauthoryear{{Gehrels}}{{Gehrels}}{2004}]{geh04}
{Gehrels} N.,  2004, in ESA Special Publication: 5th \textit{INTEGRAL} Workshop
  on the \textit{INTEGRAL} Universe. p.~777

\bibitem[\protect\citeauthoryear{HEASARC}{HEASARC}{2014}]{heasoft}
HEASARC 2014, {HEAsoft: Unified Release of FTOOLS and XANADU} (\mn@eprint
  {ascl} {1408.004})

\bibitem[\protect\citeauthoryear{{Halpern}}{{Halpern}}{2006}]{hal06}
{Halpern} J.,  2006, GRB Coordinates Network, \href
  {https://ui.adsabs.harvard.edu/abs/2006GCN..5210....1H} {5210, 1}

\bibitem[\protect\citeauthoryear{{Heger}, {Cumming}  \& {Woosley}}{{Heger}
  et~al.}{2007}]{heg07}
{Heger} A.,  {Cumming} A.,   {Woosley} S.~E.,  2007, \mn@doi [\apj]
  {10.1086/517491}, \href
  {https://ui.adsabs.harvard.edu/abs/2007ApJ...665.1311H} {665, 1311}

\bibitem[\protect\citeauthoryear{{Jahoda} et~al.}{{Jahoda}
  et~al.}{2006}]{jah06}
{Jahoda} K.,  et~al., 2006, \mn@doi [\apjs] {10.1086/500659}, \href
  {https://ui.adsabs.harvard.edu/abs/2006ApJS..163..401J} {163, 401}

\bibitem[\protect\citeauthoryear{{Jonker}, {Galloway}, {McClintock}, {Buxton},
  {Garcia}  \& {Murray}}{{Jonker} et~al.}{2004}]{jon04}
{Jonker} P.~G.,  {Galloway} D.~K.,  {McClintock} J.~E.,  {Buxton} M.,  {Garcia}
  M.,   {Murray} S.,  2004, \mn@doi [\mnras]
  {10.1111/j.1365-2966.2004.08246.x}, \href
  {https://ui.adsabs.harvard.edu/abs/2004MNRAS.354..666J} {354, 666}

\bibitem[\protect\citeauthoryear{{Jonker}, {Torres}, {Steeghs}  \&
  {Chakrabarty}}{{Jonker} et~al.}{2013}]{Jonker2013}
{Jonker} P.~G.,  {Torres} M. A.~P.,  {Steeghs} D.,   {Chakrabarty} D.,  2013,
  \mn@doi [\mnras] {10.1093/mnras/sts363}, \href
  {https://ui.adsabs.harvard.edu/abs/2013MNRAS.429..523J} {429, 523}

\bibitem[\protect\citeauthoryear{{Kaastra} \& {Bleeker}}{{Kaastra} \&
  {Bleeker}}{2016}]{kaa16}
{Kaastra} J.~S.,  {Bleeker} J.~A.~M.,  2016, \mn@doi [\aap]
  {10.1051/0004-6361/201527395}, \href
  {https://ui.adsabs.harvard.edu/abs/2016A&A...587A.151K} {587, A151}

\bibitem[\protect\citeauthoryear{{Keek}}{{Keek}}{2012}]{kee12}
{Keek} L.,  2012, \mn@doi [\apj] {10.1088/0004-637X/756/2/130}, \href
  {https://ui.adsabs.harvard.edu/abs/2012ApJ...756..130K} {756, 130}

\bibitem[\protect\citeauthoryear{{Keek} \& {Heger}}{{Keek} \&
  {Heger}}{2011}]{kee11}
{Keek} L.,  {Heger} A.,  2011, \mn@doi [\apj] {10.1088/0004-637X/743/2/189},
  \href {https://ui.adsabs.harvard.edu/abs/2011ApJ...743..189K} {743, 189}

\bibitem[\protect\citeauthoryear{{Keek}, {Ballantyne}, {Kuulkers}  \&
  {Strohmayer}}{{Keek} et~al.}{2014}]{kee14}
{Keek} L.,  {Ballantyne} D.~R.,  {Kuulkers} E.,   {Strohmayer} T.~E.,  2014,
  \mn@doi [\apj] {10.1088/0004-637X/789/2/121}, \href
  {https://ui.adsabs.harvard.edu/abs/2014ApJ...789..121K} {789, 121}

\bibitem[\protect\citeauthoryear{{Krimm} et~al.,}{{Krimm} et~al.}{2013}]{kri13}
{Krimm} H.~A.,  et~al., 2013, \mn@doi [\apjs] {10.1088/0067-0049/209/1/14},
  \href {https://ui.adsabs.harvard.edu/abs/2013ApJS..209...14K} {209, 14}

\bibitem[\protect\citeauthoryear{{Kuulkers}, {den Hartog}, {in't Zand},
  {Verbunt}, {Harris}  \& {Cocchi}}{{Kuulkers} et~al.}{2003}]{kuu03}
{Kuulkers} E.,  {den Hartog} P.~R.,  {in't Zand} J.~J.~M.,  {Verbunt} F.~W.~M.,
   {Harris} W.~E.,   {Cocchi} M.,  2003, \mn@doi [\aap]
  {10.1051/0004-6361:20021781}, \href
  {https://ui.adsabs.harvard.edu/abs/2003A&A...399..663K} {399, 663}

\bibitem[\protect\citeauthoryear{{Lewin}, {van Paradijs}  \& {Taam}}{{Lewin}
  et~al.}{1993}]{lew93}
{Lewin} W. H.~G.,  {van Paradijs} J.,   {Taam} R.~E.,  1993, \mn@doi [\ssr]
  {10.1007/BF00196124}, \href
  {https://ui.adsabs.harvard.edu/abs/1993SSRv...62..223L} {62, 223}

\bibitem[\protect\citeauthoryear{{Lund} et~al.,}{{Lund} et~al.}{2003}]{jemx}
{Lund} N.,  et~al., 2003, \mn@doi [\aap] {10.1051/0004-6361:20031358}, \href
  {https://ui.adsabs.harvard.edu/abs/2003A&A...411L.231L} {411, L231}

\bibitem[\protect\citeauthoryear{{Mancuso}, {Altamirano}, {M{\'e}ndez}, {Lyu}
  \& {Combi}}{{Mancuso} et~al.}{2021}]{man21}
{Mancuso} G.~C.,  {Altamirano} D.,  {M{\'e}ndez} M.,  {Lyu} M.,   {Combi}
  J.~A.,  2021, \mn@doi [\mnras] {10.1093/mnras/stab159}, \href
  {https://ui.adsabs.harvard.edu/abs/2021MNRAS.502.1856M} {502, 1856}

\bibitem[\protect\citeauthoryear{{Marino} et~al.,}{{Marino}
  et~al.}{2022}]{mar22}
{Marino} A.,  et~al., 2022, \mn@doi [\mnras] {10.1093/mnras/stac2038}, \href
  {https://ui.adsabs.harvard.edu/abs/2022MNRAS.515.3838M} {515, 3838}

\bibitem[\protect\citeauthoryear{{Markwardt} \& {Strohmayer}}{{Markwardt} \&
  {Strohmayer}}{2010}]{mar10}
{Markwardt} C.~B.,  {Strohmayer} T.~E.,  2010, \mn@doi [\apjl]
  {10.1088/2041-8205/717/2/L149}, \href
  {https://ui.adsabs.harvard.edu/abs/2010ApJ...717L.149M} {717, L149}

\bibitem[\protect\citeauthoryear{{Mereminskiy}, {Grebenev}, {Lutovinov},
  {Krivonos}  \& {Kuulkers}}{{Mereminskiy} et~al.}{2021}]{2021ATel14427}
{Mereminskiy} I.~A.,  {Grebenev} S.~A.,  {Lutovinov} A.~A.,  {Krivonos} R.~A.,
   {Kuulkers} E.,  2021, The Astronomer's Telegram, \href
  {https://ui.adsabs.harvard.edu/abs/2021ATel14427....1M} {14427, 1}

\bibitem[\protect\citeauthoryear{{Mi{\v{s}}kovi{\v{c}}ov{\'a}}
  et~al.,}{{Mi{\v{s}}kovi{\v{c}}ov{\'a}} et~al.}{2016}]{mis16}
{Mi{\v{s}}kovi{\v{c}}ov{\'a}} I.,  et~al., 2016, \mn@doi [\aap]
  {10.1051/0004-6361/201322490}, \href
  {https://ui.adsabs.harvard.edu/abs/2016A&A...590A.114M} {590, A114}

\bibitem[\protect\citeauthoryear{{Ootes}, {Watts}, {Galloway}  \&
  {Wijnands}}{{Ootes} et~al.}{2017}]{oot17}
{Ootes} L.~S.,  {Watts} A.~L.,  {Galloway} D.~K.,   {Wijnands} R.,  2017,
  \mn@doi [\apj] {10.3847/1538-4357/834/1/21}, \href
  {https://ui.adsabs.harvard.edu/abs/2017ApJ...834...21O} {834, 21}

\bibitem[\protect\citeauthoryear{{Palmer} et~al.,}{{Palmer}
  et~al.}{2006}]{pal06}
{Palmer} D.,  et~al., 2006, GRB Coordinates Network, \href
  {https://ui.adsabs.harvard.edu/abs/2006GCN..5208....1P} {5208, 1}

\bibitem[\protect\citeauthoryear{{Patruno} \& {Watts}}{{Patruno} \&
  {Watts}}{2021}]{pat21}
{Patruno} A.,  {Watts} A.~L.,  2021, \mn@doi [Astrophysics and Space Science
  Library] {10.1007/978-3-662-62110-3\_4}, \href
  {https://ui.adsabs.harvard.edu/abs/2021ASSL..461..143P} {461, 143}

\bibitem[\protect\citeauthoryear{{Pavan} et~al.,}{{Pavan} et~al.}{2010}]{pav10}
{Pavan} L.,  et~al., 2010, The Astronomer's Telegram, \href
  {https://ui.adsabs.harvard.edu/abs/2010ATel.2548....1P} {2548, 1}

\bibitem[\protect\citeauthoryear{{Ponti}, {Mu{\~n}oz-Darias}  \&
  {Fender}}{{Ponti} et~al.}{2014}]{pon14}
{Ponti} G.,  {Mu{\~n}oz-Darias} T.,   {Fender} R.~P.,  2014, \mn@doi [\mnras]
  {10.1093/mnras/stu1742}, \href
  {https://ui.adsabs.harvard.edu/abs/2014MNRAS.444.1829P} {444, 1829}

\bibitem[\protect\citeauthoryear{{Remillard} et~al.,}{{Remillard}
  et~al.}{2022}]{nibackgen}
{Remillard} R.~A.,  et~al., 2022, \mn@doi [\aj] {10.3847/1538-3881/ac4ae6},
  \href {https://ui.adsabs.harvard.edu/abs/2022AJ....163..130R} {163, 130}

\bibitem[\protect\citeauthoryear{{Revnivtsev}, {Churazov}, {Gilfanov}  \&
  {Sunyaev}}{{Revnivtsev} et~al.}{2001}]{rev01}
{Revnivtsev} M.,  {Churazov} E.,  {Gilfanov} M.,   {Sunyaev} R.,  2001, \mn@doi
  [\aap] {10.1051/0004-6361:20010434}, \href
  {https://ui.adsabs.harvard.edu/abs/2001A&A...372..138R} {372, 138}

\bibitem[\protect\citeauthoryear{{Sanna} et~al.,}{{Sanna} et~al.}{2022}]{san22}
{Sanna} A.,  et~al., 2022, \mn@doi [\mnras] {10.1093/mnras/stac1611}, \href
  {https://ui.adsabs.harvard.edu/abs/2022MNRAS.514.4385S} {514, 4385}

\bibitem[\protect\citeauthoryear{{Schady}, {Beardmore}, {Marshall}, {Palmer},
  {Rol}  \& {Sato}}{{Schady} et~al.}{2006}]{sch06}
{Schady} P.,  {Beardmore} A.~P.,  {Marshall} F.~E.,  {Palmer} D.~M.,  {Rol} E.,
    {Sato} G.,  2006, GRB Coordinates Network, \href
  {https://ui.adsabs.harvard.edu/abs/2006GCN..5200....1S} {5200, 1}

\bibitem[\protect\citeauthoryear{{Strohmayer} \& {Bildsten}}{{Strohmayer} \&
  {Bildsten}}{2006}]{str03}
{Strohmayer} T.,  {Bildsten} L.,  2006, in , Vol.~39, Compact stellar X-ray
  sources.
pp 113--156

\bibitem[\protect\citeauthoryear{{Strohmayer} \& {Markwardt}}{{Strohmayer} \&
  {Markwardt}}{2010}]{str10}
{Strohmayer} T.~E.,  {Markwardt} C.~B.,  2010, The Astronomer's Telegram, \href
  {https://ui.adsabs.harvard.edu/abs/2010ATel.2569....1S} {2569, 1}

\bibitem[\protect\citeauthoryear{{Str{\"u}der} et~al.,}{{Str{\"u}der}
  et~al.}{2001}]{xmm}
{Str{\"u}der} L.,  et~al., 2001, \mn@doi [\aap] {10.1051/0004-6361:20000066},
  \href {https://ui.adsabs.harvard.edu/abs/2001A&A...365L..18S} {365, L18}

\bibitem[\protect\citeauthoryear{{Tauris} \& {van den Heuvel}}{{Tauris} \& {van
  den Heuvel}}{2006}]{tau06}
{Tauris} T.~M.,  {van den Heuvel} E.~P.~J.,  2006, {Formation and evolution of
  compact stellar X-ray sources}.
Cambridge University Press, Cambridge, pp 623--665

\bibitem[\protect\citeauthoryear{{Verner}, {Ferland}, {Korista}  \&
  {Yakovlev}}{{Verner} et~al.}{1996}]{ver96}
{Verner} D.~A.,  {Ferland} G.~J.,  {Korista} K.~T.,   {Yakovlev} D.~G.,  1996,
  \mn@doi [\apj] {10.1086/177435}, \href
  {https://ui.adsabs.harvard.edu/abs/1996ApJ...465..487V} {465, 487}

\bibitem[\protect\citeauthoryear{{Wijnands}, {Rol}, {Cackett}, {Starling}  \&
  {Remillard}}{{Wijnands} et~al.}{2009}]{wij09}
{Wijnands} R.,  {Rol} E.,  {Cackett} E.,  {Starling} R.~L.~C.,   {Remillard}
  R.~A.,  2009, \mn@doi [\mnras] {10.1111/j.1365-2966.2008.14175.x}, \href
  {https://ui.adsabs.harvard.edu/abs/2009MNRAS.393..126W} {393, 126}

\bibitem[\protect\citeauthoryear{{Wilms}, {Allen}  \& {McCray}}{{Wilms}
  et~al.}{2000}]{wil00}
{Wilms} J.,  {Allen} A.,   {McCray} R.,  2000, \mn@doi [\apj] {10.1086/317016},
  \href {https://ui.adsabs.harvard.edu/abs/2000ApJ...542..914W} {542, 914}

\bibitem[\protect\citeauthoryear{{Worpel}, {Galloway}  \& {Price}}{{Worpel}
  et~al.}{2013}]{wor13}
{Worpel} H.,  {Galloway} D.~K.,   {Price} D.~J.,  2013, \mn@doi [\apj]
  {10.1088/0004-637X/772/2/94}, \href
  {https://ui.adsabs.harvard.edu/abs/2013ApJ...772...94W} {772, 94}

\bibitem[\protect\citeauthoryear{{Worpel}, {Galloway}  \& {Price}}{{Worpel}
  et~al.}{2015}]{wor15}
{Worpel} H.,  {Galloway} D.~K.,   {Price} D.~J.,  2015, \mn@doi [\apj]
  {10.1088/0004-637X/801/1/60}, \href
  {https://ui.adsabs.harvard.edu/abs/2015ApJ...801...60W} {801, 60}

\makeatother
\end{thebibliography}



\appendix

\section{Varying \NH\ method in X-ray burst time-resolved spectroscopy}
\label{sec:app}

\begin{figure}
    \centering
    \includegraphics[width=\linewidth, trim=0 0 0 0, clip]{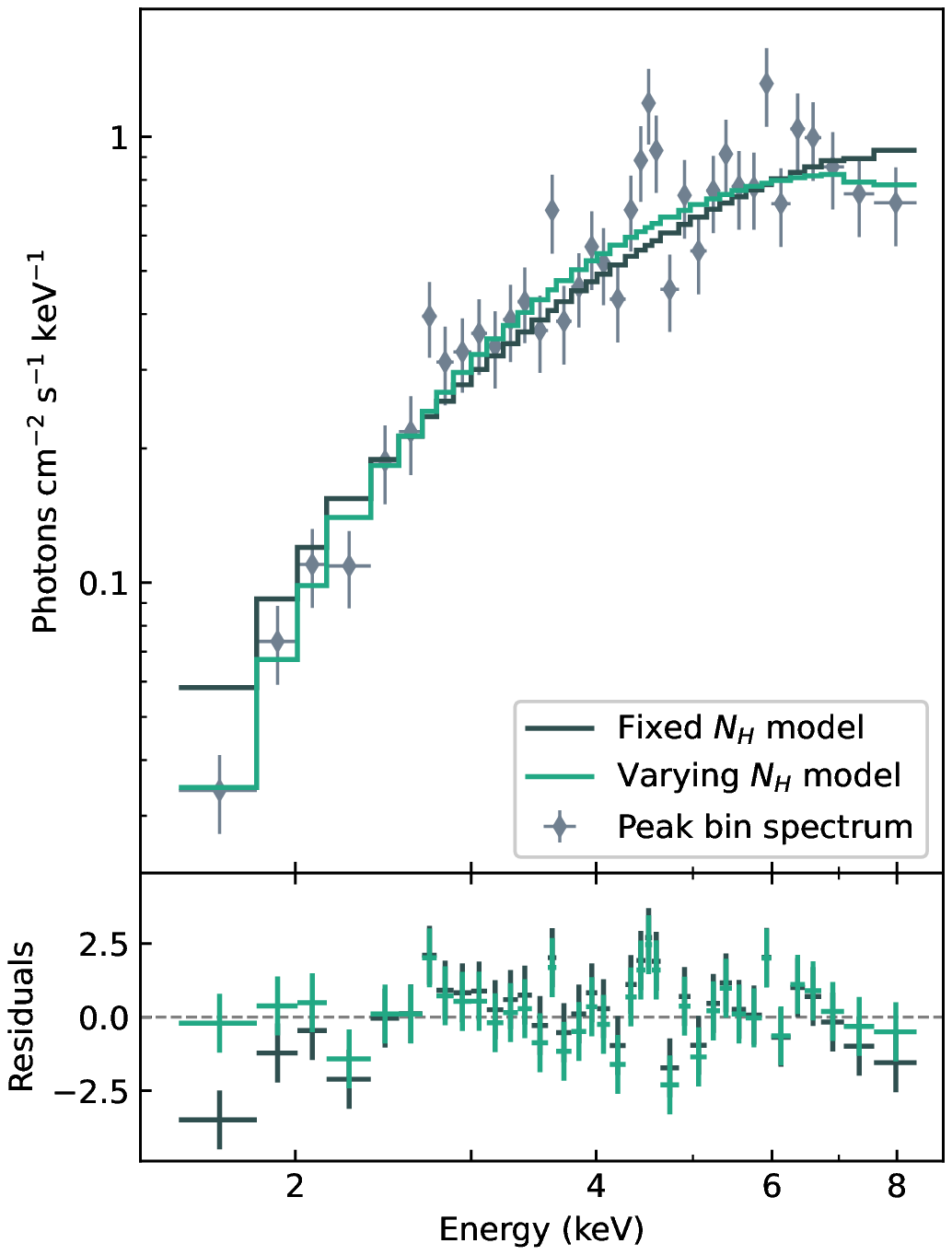}
    
    \caption{Burst 4 spectrum, models, and residuals from the peak of the burst (bin 4 of the time-resolved spectroscopy). Top: unfolded burst spectrum (\texttt{eufspec} command in \texttt{Xspec}), and best fit models for the fixed \NH\ method and varying \NH\ method. Bottom: residuals from each of the models (\texttt{delchi} command in \texttt{Xspec}).}
    
    \label{fig:app_spec}
\end{figure}

\begin{figure*}
    \centering
    \includegraphics[width=\linewidth, trim=0 0 0 0, clip]{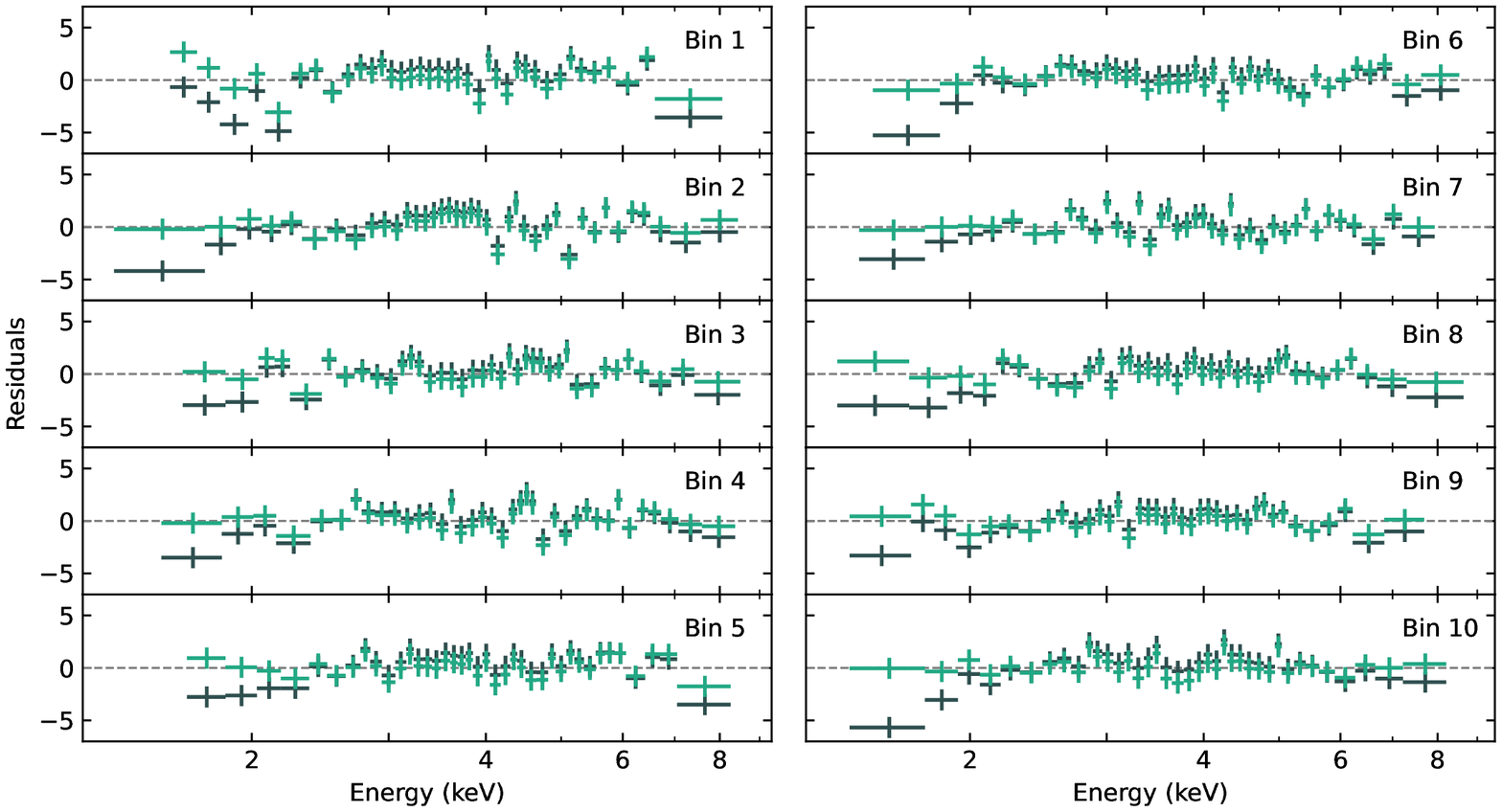}
    
    \caption{Residuals from the first 10 spectra used in the time-resolved spectroscopy of burst 4, where the spectra have been fitted with the fixed and varying \NH\ methods (\texttt{delchi} command in \texttt{Xspec}). The colour coding is as presented in Figure \ref{fig:app_spec} (dark grey for fixed \NH\ method and teal for varying \NH\ method).}

    \label{fig:app_res}
\end{figure*}

In this appendix, we expand upon our investigations for the X-ray burst time-resolved spectroscopy presented in Section \ref{sec:trs}. We found that a single black body model,

\begin{quote}
    \texttt{tbabs $\times$ (bbodyrad + powerlaw)}
\end{quote}

\noindent with \NH\ and powerlaw components fixed to the best fit parameters from the persistent emission fits (see Table \ref{tab:pe}) resulted in large residuals. However, allowing \NH\ to vary freely reduced these residuals. In Figure \ref{fig:app_spec} we show an example spectrum taken from the peak bin of B4. We can see the model best fits for both the fixed and varying \NH\ methods, and the residuals. Figure \ref{fig:app_res} shows the residuals from the first 10 bins of B4, which covers 29\,s since the burst start time. The difference in residuals between the two models shows that the varying \NH\ model better fits the data, particularly in the lower and higher energy ranges of the spectra.  

\begin{figure*}
    \centering
    \includegraphics[width=\linewidth, trim=0 0 0 0, clip]{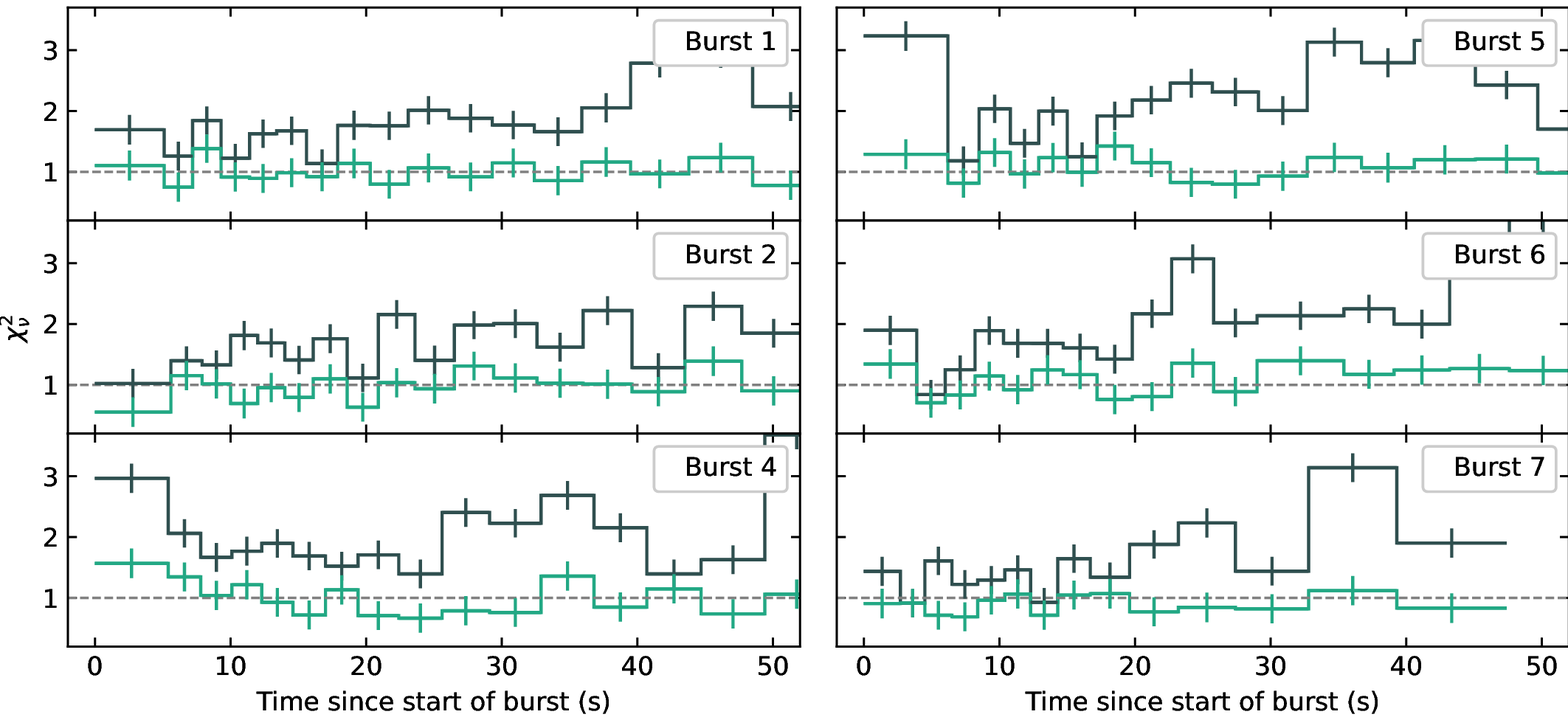}
    
    \caption{Reduced \chis\ panels from the time-resolved spectroscopy of each burst, where the spectra have been fitted using the fixed and varying \NH\ methods. The colour coding is as presented in Figure \ref{fig:app_spec} (dark grey for fixed \NH\ method and teal for varying \NH\ method).}
    
    \label{fig:app_chi2}
\end{figure*}

Reduced \chis\ panels from implementing both the fixed and varying \NH\ methods in the time-resolved spectroscopy of all analysed bursts can be seen in Figure \ref{fig:app_chi2}. In each case, the reduced \chis\ values indicate that the fixed \NH\ method did not produce suitable fits, and allowing \NH\ to vary improved the fits.

We note that implementing the $f_a$-method \citep{wor13,wor15} to scale the persistent emission only produced reasonable fits when $f_a$ was allowed to be negative, which is not a physically motivated option. 

\bsp	
\label{lastpage}
\end{document}